\documentclass[camera,letterpaper,nomarginnotes,nonarrowgutter]{jpaper}
    
\usepackage{amsmath,amssymb,amsfonts}
\usepackage{algorithm}
\usepackage[noend]{algpseudocode}
\usepackage{algorithmicx}
\usepackage{graphicx}
\usepackage{textcomp}

\usepackage{makecell}
\usepackage{array}
\usepackage{booktabs}

\usepackage[sort,compress]{cite}

\usepackage{hyperref}
  \hypersetup{%
    unicode=true,%
    pdfstartview={FitH},%
    colorlinks=true,%
    citecolor=blue,%
    filecolor=blue,%
    linkcolor=black,%
    urlcolor=black}
    
\usepackage{soul}
\usepackage{xspace}
\usepackage{multirow}
\usepackage{tabularx}
\usepackage{tikz}
\usepackage[bottom]{footmisc}
\newcommand*\circled[1]{\tikz[baseline=(char.base)]{\node[shape=circle,fill,inner sep=0.5pt] (char) {\textcolor{white}{#1}};}}
\usepackage[most]{tcolorbox}

\usepackage{fancyhdr}
\usepackage{mathtools}

\usepackage{subcaption}
\usepackage{booktabs}

\usepackage{pgfplots}
\pgfplotsset{compat=1.17} %

\usepackage{stfloats}

\usepackage{enumitem}

\usepackage{mwe}

\definecolor{lavenderr}{rgb}{0.71, 0.49, 0.86}

\usepackage[most]{tcolorbox}   %
\tcbset{
  observebox/.style={
    colframe=lavenderr!90!black,      %
    colback=lavenderr!15,              %
    boxrule=1pt,                   %
    arc=1mm,                       %
    left=0.1mm, right=0.1mm,       %
    top=0.1mm, bottom=0.1mm,       %
    fonttitle=\bfseries,           %
    before skip=6pt, after skip=6pt
  }
}

\usepackage{graphicx}

\usepackage{tikzducks}

\usepackage[en-GB]{datetime2} 
\DTMsetup{
    showzone=false,   
    showseconds=false 
}

\definecolor{darkspringgreen}{rgb}{0.09, 0.45, 0.27}
\definecolor{denim}{rgb}{0.08, 0.38, 0.74}
\definecolor{darkolivegreen}{rgb}{0.33, 0.42, 0.18}
\definecolor{tangerine}{rgb}{0.95, 0.52, 0.0}
\definecolor{mahogany}{rgb}{0.75, 0.25, 0.0}
\definecolor{coolblack}{rgb}{0.0, 0.22, 0.44}
\definecolor{darkpink}{rgb}{0.91, 0.35, 0.6}
\definecolor{darkblue}{rgb}{0.0, 0.0, 0.67}
\definecolor{melon}{rgb}{0.97, 0.69, 0.67}

\definecolor{seagreen}{rgb}{0.18, 0.55, 0.34}

\definecolor{pred}{rgb}{0.7843, 0.0039, 0.3137} 

\definecolor{darkpink}{rgb}{0.88, 0.28, 0.54}
\definecolor{forestgreen}{rgb}{0.0, 0.27, 0.13}
\definecolor{amber}{rgb}{1.0, 0.49, 0.0}

\newcommand{\inum}[1]{(\textit{#1})\xspace}

\newcommand{\sect}[1]{{§#1}\xspace} %
\newcommand{\head}[1]{{\vspace{2pt}\noindent\textbf{#1.}\xspace}} %
\newcommand{\fig}[1]{{Fig.~#1}\xspace} %

\newcommand\ssdm{\texttt{SSD-G4}\xspace}
\newcommand\ssdh{\texttt{SSD-G5}\xspace}
\newcommand\dram{\texttt{\omciv{DRAM}}\xspace}

\newcolumntype{Y}{>{\centering\arraybackslash}X}
\usepackage{tikz}

\newcommand{\squishlist}{
 \begin{list}{$\circ$}
  { \setlength{\itemsep}{0pt}
     \setlength{\parsep}{0pt}
     \setlength{\topsep}{3pt}
     \setlength{\partopsep}{0pt}
     \setlength{\leftmargin}{1em}
     \setlength{\labelwidth}{1em}
     \setlength{\labelsep}{0.5em} } }

\newcommand{\squishend}{
  \end{list}  }

\makeatletter
\g@addto@macro{\normalsize}{%
  \setlength{\abovedisplayskip}{4pt plus 0.5pt minus 1pt}
  \setlength{\belowdisplayskip}{4pt plus 0.5pt minus 1pt}
  \setlength{\abovedisplayshortskip}{0pt}
  \setlength{\belowdisplayshortskip}{0pt}
  \setlength{\intextsep}{3pt plus 1pt minus 1pt}
  \setlength{\textfloatsep}{7pt plus 1pt minus 1pt}
  \setlength{\skip\footins}{5pt plus 1pt minus 0pt}}
\makeatother

\usepackage{blindtext,graphicx}
\usepackage[absolute]{textpos}
\usepackage{datetime}

\definecolor{seagreen}{rgb}{0.18, 0.55, 0.34}
\definecolor{ballblue}{rgb}{0.13, 0.67, 0.8}

\definecolor{darkgreen}{rgb}{0.0, 0.44, 0.34}

\sloppypar

\definecolor{dollarbill}{rgb}{0.52, 0.73, 0.4}

\usetikzlibrary{patterns}
\usepackage[precision=2, unit=mm]{lengthconvert}

\usepackage[colorinlistoftodos,prependcaption,textsize=scriptsize]{todonotes}
\usepackage{marginnote} 

\definecolor{cyan(process)}{rgb}{0.0, 0.62, 0.82}

\newcommand\proposal{GRAINS\xspace}

\definecolor{cadmiumgreen}{rgb}{0.0, 0.50, 0.29}

\newcommand\revref[1]{\hyperref[rev:#1]{#1}}

\definecolor{raspberry}{rgb}{0.89, 0.04, 0.36}

\definecolor{awesome}{rgb}{1.0, 0.13, 0.32}
\definecolor{cardinal}{rgb}{0.77, 0.12, 0.23}
\definecolor{cadet}{rgb}{0.33, 0.41, 0.47}
\definecolor{celadon}{rgb}{0.67, 0.88, 0.69}
\definecolor{persianblue}{rgb}{0.11, 0.22, 0.73}
\definecolor{ultramarine}{rgb}{0.07, 0.04, 0.56}
\definecolor{warmblack}{rgb}{0.0, 0.3, 0.3}

\definecolor{terracotta}{rgb}{0.89, 0.45, 0.36}

\definecolor{forestgreen(web)}{rgb}{0.13, 0.55, 0.13}

\renewcommand\dram{\texttt{No-I/O}\xspace}

\newcommand\strings{\textsf{Strings}\xspace}
\newcommand\offsets{\textsf{Offsets}\xspace}
\newcommand\sizes{\textsf{Sizes}\xspace}
\newcommand\colors{\textsf{Colors}\xspace}

\renewcommand{\figurename}{Fig.}

\definecolor{cardinal}{rgb}{0.77, 0.12, 0.23}
\definecolor{deeppink}{rgb}{1.0, 0.08, 0.58}
\definecolor{brightpink}{rgb}{1.0, 0.0, 0.5}
\definecolor{electricviolet}{rgb}{0.56, 0.0, 1.0}
\definecolor{brandeisblue}{rgb}{0.0, 0.44, 1.0}
\definecolor{carminered}{rgb}{1.0, 0.0, 0.22}
\newcommand\asp[1]{{\color{black}{#1}}} %

\usepackage{setspace}
\definecolor{acolor}{rgb}{0.0, 0.5, 1.0}
\definecolor{bcolor}{rgb}{0.54, 0.17, 0.89}
\definecolor{ccolor}{rgb}{0.4, 0.69, 0.2}
\definecolor{dcolor}{rgb}{0.92, 0.41, 0.12}
\definecolor{ecolor}{rgb}{0.6, 0.0, 0.156}
\definecolor{fcolor}{rgb}{0.106, 0.620, 0.467}

\newcommand\irevminor[1]{{\color{coolblack}{#1}}}

\newcommand{\citegraphacc}{milanese2019microbial,salzberg2016next,gihawi2023major,pockrandt2022metagenomic,berger2023navigating,Ackelsberg2015,Nasko2018,meyer2021critical,alser2020technology,alser2022molecules}

\definecolor{dogwoodrose}{rgb}{0.84, 0.09, 0.41}

\newcommand\omcr[1]{{\color{dogwoodrose}{#1}}}

\newcommand\omii[1]{{\color{orange}{#1}}}

\newcommand\omiiq[1]{\todo[linecolor=orange,backgroundcolor=orange!10,bordercolor=orange]{\textcolor{orange}{\textit{\scriptsize #1}}}}

\newcommand\cutcr[1]{\todo[linecolor=red,backgroundcolor=red!10,bordercolor=red]{\textcolor{red}{\textit{\scriptsize #1}}}}

\newif\ifsubmit

\submittrue

\definecolor{darkpastelgreen}{rgb}{0.01, 0.75, 0.24}
\definecolor{darkcyan}{rgb}{0.0, 0.55, 0.55}
\definecolor{ferngreen}{rgb}{0.31, 0.47, 0.26}
\definecolor{indiagreen}{rgb}{0.07, 0.53, 0.03}
\definecolor{mint}{rgb}{0.24, 0.71, 0.54}
\definecolor{oliv}{rgb}{0.42, 0.56, 0.14}

\newcommand\omiii[1]{{\color{darkpastelgreen}{#1}}}
\newcommand\omiiiq[1]{\todo[linecolor=darkpastelgreen,backgroundcolor=darkpastelgreen!10,bordercolor=darkpastelgreen]{\textcolor{darkpastelgreen}{\textit{\scriptsize #1}}}}

\ifsubmit
\renewcommand\todo[1]{}
\renewcommand\cutcr[1]{}
\renewcommand\omiiq[1]{}
\renewcommand\omiiiq[1]{}
\renewcommand\omii[1]{{\color{black}{#1}}}
\renewcommand\omiii[1]{{\color{black}{#1}}}
\renewcommand\omcr[1]{{\color{black}{#1}}}

\else
\fi

\newcommand{\citepersonalized}{alkan_personalized_2009,lightbody_review_2019,morganti_next_2019,branco_bioinformatics_2021,quazi_artificial_2022,aronson_building_2015,f_lochel_comparative_2020,papadopoulou_application_2023,tafazoli_applying_2021,gambardella_personalized_2020,leary_development_2010,hamburg_margaret_a_path_2010,van_der_lee_technologies_2020,moon_precision_2022,mohan_profiling_2020,chung_rapid_2020,bielinski_preemptive_2014,ho_enabling_2020,hussen_emerging_2022,russell_pharmacogenomics_2021,verma_nanopore_2024,clark2019diagnosis,farnaes2018rapid,sweeney2021rapid,flores2013p4,ginsburg2009genomic,chin2011cancer,Ashley2016}

\newcommand{\citeoutbreakrapid}{dunn_squigglefilter_2021,bertelli_rapid_2013,arias_rapid_2016,comin_investigation_2020,Quick2016}

\newcommand{\citeoutbreak}{\citeoutbreakrapid,robinson_genomics_2013,fournier_clinical_2014,koser_routine_2012,eloit_diagnosis_2014,gardy_jennifer_l_whole-genome_2011,taylor_angela_j_characterization_2015,quainoo_scott_whole-genome_2017,goldberg_brittany_making_2015,besser_interpretation_2019,li_application_2021,deng_integrated_2021,kwong_whole_2015,deurenberg_application_2017,tang_infection_2017,croucher_application_2015,
bloom2021massively,yelagandula2021multiplexed,le2013selected,nikolayevskyy2016whole,qiu2015whole,gilchrist2015whole}

\newcommand{\citeagriculture}{the_arabidopsis_genome_initiative_analysis_2000,zhu_applications_2020,choi_nanopore_2020,stevens_sequence_2016,campos_high_2021,gao_genome_2021,van_dijk_machine_2021,sun_twenty_2022,kim_application_2020,thudi_genomic_2021,michael_building_2020,shen_omics-based_2022,shahroodi2022demeter,
prasad2021soil,Mascher2024,Schreiber2024}

\newcommand{\citeevolution}{kanehisa_toward_2019,qing_whole_2022,wittkopp_cis-regulatory_2012,romero_comparative_2012,wang_population_2020,hill_molecular_2021,vaishnav_evolution_2022,zhang_haplotype-resolved_2021,fay_evaluating_2008,kanzi_next_2020,wray_evolution_2003,wu_one_2021,signor_evolution_2018,whitehead_variation_2006,coolon_tempo_2014,
ellegren2014genome,Prado-Martinez2013,Prohaska2019human}

\newcommand{\citecancer}{lawrence_mutational_2013,vogelstein_cancer_2013,ramskold_full-length_2012,baslan_unravelling_2017,shapiro_single-cell_2013,sakamoto_new_2020,jia_high-throughput_2022,lawson_tumour_2018,liu_mrna-based_2023,van_de_sande_applications_2023,chakravarty_clinical_2021,cortes-ciriano_computational_2022,deveson_evaluating_2021,xiao_toward_2021,bolton_cancer_2020,szustakowski_advancing_2021,navin_future_2011,hong_rna_2020,lei_applications_2021,han_single-cell_2022,federici_variants_2020,zhang_singlecell_2021,ren_understanding_2018,tian_cicero_2020,malone_molecular_2020,tang_single-cell_2019,ellsworth_single-cell_2017,zhong_application_2021,stadler_therapeutic_2021,tan_targeted_2022,degasperi_substitution_2022,xu_single-cell_2022,horak_comprehensive_2021,zhang_single-cell_2016,bruno_next_2020,de_luca_fgfr_2020,waarts_targeting_2022,lim_advancing_2020,colomer_when_2020,saadatpour_single-cell_2015,dizman_sequencing_2020,buzdin_rna_2020,xiao_tumor_2021,nandwani_lncrnas_2021,marchetti_error-corrected_2023,chen_next-generation_2021,navin_first_2015}

\newcommand{\citehwoptimization}{doblas2025smx,mutlu2023accelerating,alser2022molecules,lou2020helix,lou2018brawl,shahroodi2023swordfish,markus2020benchmarking,subramaniyan2021accelerated,huangfu2018radar,khatamifard2021genvom,gupta2019rapid,li2021pim,angizi2019aligns,zokaee2018aligner,turakhia2018darwin,fujiki2018genax,madhavan2014race,cheng2018bitmapper2,houtgast2018hardware,houtgast2017efficient,zeni2020logan,ahmed2019gasal2,nishimura2017accelerating,de2016cudalign,liu2015gswabe,liu2013cudasw++,liu2009cudasw++,liu2010cudasw++,wilton2015arioc,goyal2017ultra,chen2016spark,chen2014accelerating,chen2021high,fujiki2020seedex,banerjee2018asap,fei2018fpgasw,waidyasooriya2015hardware,chen2015novel,rucci2018swifold,haghi2021fpga,li2021pipebsw,ham2020genesis,ham2021accelerating,wu2019fpga,cali2020genasm,Zhang_2023_alignerD,soysal2025mars,kim2018grim,kaplan2020bioseal,mao2022genpip,dphls2026,wang20202,Walia2024talco,sadasivan2024genomic,Turakhia2025toward,Turakhia2019darwinwga,simon2026pim}
\newcommand{\citemghwoptimization}{jia2011metabing,kobus2021metacache,wang2023gpmeta,kobus2017accelerating,Su2012,su2013gpumetastorms,Yano2014,saavedra2020mining,zhang2023genomix,cervi2022metagenomic,wu2021sieve,shahroodi2022krakenonmem,shahroodi2022demeter,dashcam23micro,hanhan2022edam,zou2022biohd,dunn_squigglefilter_2021,shih2023efficient}

\newcommand{\citegraphhwoptimization}{Cali2022SeGraM,Zhang2024Harp,Zeng2024asgdp,Shen2024128parallel,Li2024,Mandal2020,Varma2013,Awan2021,Feng2021,Zhang2025,kim2025nmp,Huang2023meg2,Angizi2020Panda,Qiu2017,Zhou2021,Sarkar2021,Varma2017,Varma2016,Goswami2018,Galanos2021,Angizi2020,Sinha2022,Meng2014,Hu2016,Chen2023,Natarajan2018,Ren2018}

\begin{document}

\title{\LARGE\proposal: Storage-Aware Algorithm-Architecture Co-Design\\ Enabling High-Performance and Low-Cost\\ Graph-Based
Genome Analysis}

\def\iscacameraready{} %
\newcommand{\hpcapubid}{0000--0000/00\$00.00}

\newcommand\iscaauthors{
Nika Mansouri Ghiasi$^1$ \hspace{0.5em} Harun Mustafa$^{2,1}$  \hspace{0.5em} Talu Güloglu$^1$ \hspace{0.5em} Rakesh Nadig$^1$\\
Konstantina Koliogeorgi$^1$ \hspace{0.5em} Susana Rebolledo Ruiz$^{1,3}$ \hspace{0.5em} Marc Rautmann$^1$ \hspace{0.5em} Furkan Eris$^1$\\
Mohammad Sadrosadati$^1$ \hspace{0.5em} Jisung Park$^4$ \hspace{0.5em} Onur Mutlu$^1$\\\\
}

\newcommand\iscaaffiliation{
\hspace{-2.5em} $^1$ETH Zürich \hspace{0.5em} $^2$Johns Hopkins University \hspace{0.5em} $^3$University of Cantabria \hspace{0.5em} $^4$POSTECH
}

\author{
\iscaauthors{}
\iscaaffiliation{}
}

\pagestyle{fancy}
\fancyhf{} 
\renewcommand{\headrulewidth}{0pt}

\ifsubmit
\else
\fancyhead[C]{\textcolor{blue}{\textit{V2.1 -- \DTMnow{} UTC}}}
\fi

\renewcommand{\headrulewidth}{0pt}

\maketitle

\newcommand{\iscaheight}{0mm}
\ifdefined\eaopen
\renewcommand{\iscaheight}{12mm}
\fi

\setcounter{page}{1}

\begin{abstract}
\asp{Graph-based representations of genome sequences have emerged as a powerful approach for representing massive genomic databases in \omcr{an} expressive and efficient \omcr{way}. Compared to traditional, linear genome sequences, genome graphs enable more accurate and efficient genome analyses, particularly in complex, population-scale settings (e.g., public health, precision medicine, and agriculture).}    
Despite their benefits, analysis on large-scale genome graphs incurs significant data movement overhead from the storage system due to accessing large amounts of low-reuse data. Processing data directly inside the storage device, where data originally resides, can be a fundamental solution for mitigating this overhead. However, none of the existing tools for graph-based genome analysis can be efficiently used inside the storage system due to the limited internal hardware resources in modern SSDs. \asp{At the same time, prior storage-centric systems developed for \inum{i}~traditional, linear non-graph-based genome analysis or \inum{ii}~conventional, non-genomic graph analysis are \omcr{not suitable} for the unique data structures and access patterns of graph-based genome analysis.}

We propose \proposal, the \emph{first} system for analysis with large-scale \underline{\textbf{g}}enome g\underline{\textbf{ra}}phs \underline{\textbf{in}} \underline{\textbf{s}}torage. 
Through our detailed examination of typical analysis pipelines \omcr{that operate} on genome graphs, we \asp{perform storage-aware algorithm-architecture co-design to} \inum{i}~make the \omcr{graph-based genome analysis} pipelines more storage-friendly and \inum{ii}~further improve performance, energy-efficiency, and cost via in-storage and in-flash processing. 
\asp{\proposal{}'s co-design is based on three key aspects.
First, we propose a new batching technique and execution flow, based on unique features of genome graphs, that reduces the number of random accesses \omcr{to graph nodes}. Second, via in-flash and in-storage processing, we avoid transferring low-reuse or unused flash pages, preventing SSD channel and external I/O bandwidth waste.
Third, to leverage the full parallelism \omcr{of flash dies} during in-flash processing, we design an effective, yet lightweight, scheduling technique, enabled by re-purposing the existing SSD structures.}
\proposal's design is versatile and flexible as it supports key operations on genome graphs and can be integrated in various \omcr{analysis} pipelines. \proposal provides 2.7$\times$–47.8$\times$ speedup \omcr{(4.4$\times$–31.6$\times$ energy reduction)} over the state-of-the-art software baselines, and 1.5$\times$–17.0$\times$ speedup \omcr{(3.1$\times$–20.7$\times$ energy reduction)} over a hardware-accelerated baseline.

\end{abstract}

\vspace{-0.5em}
\section{Introduction}
\label{sec:intro}

Genome sequence analysis, 
which examines the genomic information of living organisms
and other biological entities, is fundamental to many critical fields, including personalized medicine\omii{~\cite{\citepersonalized}}, disease outbreak tracing\omii{~\cite{\citeoutbreak}}, \omii{cancer research~\cite{\citecancer}}, maintenance of food resource safety~\cite{e002244,TONG2021130}, \omii{agriculture~\cite{\citeagriculture}, scientific discovery~\cite{urbanek2018degradation,edgar2022petabase,paoli2022biosynthetic}, biodiversity conservation~\cite{Hogg2024,lewin2018earth},  \omiii{evolutionary biology~\cite{\citeevolution}, and antimicrobial resistance surveillance~\cite{danko2021global,didelot2012transforming,marini2022towards}.}} 
To enable the computational analysis of an organism's genomic information (typically encoded in DNA or RNA that has been reverse-transcribed into DNA~\cite{Houldcroft2017,Jansz2024viral}), a process called \emph{sequencing} converts the information of DNA molecules in the organism's biological sample to digital data. \asp{Since current sequencing technologies cannot process a DNA molecule as a whole, a sequencer generates randomly sampled, inexact fragments of genomic information, called \emph{reads}.
Common genome analysis workflows then involve querying reads 
from a sample
against a single \emph{reference} genome or large-scale databases (e.g., containing reference genomes from many species or previously sequenced samples) to determine species or genomic features present in the sample~\cite{karasikov2020metagraph,alanko2023themisto,Muggli2017SuccinctGraphs,fan2023fulgor,piro2020ganon,mangul2016reference} and/or to find their differences from genomes in the database~\cite{aganezov2022complete,Groza2024}. These queries either involve k-mer set
lookups (where subsequences of length $k$, i.e., k-mers, in a read set are matched against the database) or read mapping (where each read is analyzed individually).}
The importance of genome analysis, along with rapid improvements in sequencing technology (i.e., reduced costs and increased throughput~\cite{berger2023navigating,katz2021sra}), has led to 
its increasing adoption~\cite{clark2019diagnosis,farnaes2018rapid,sweeney2021rapid,ginsburg2009genomic,chin2011cancer,Ashley2016,bloom2021massively,gilchrist2015whole}.
\asp{This growth has, in turn, motivated a large body of work (e.g.,\omii{~\cite{\citehwoptimization,\citemghwoptimization,\citegraphhwoptimization}}) on accelerating these analyses to keep pace with rapidly growing sequencing throughput and data volumes}.

\noindent\textbf{Genome Graphs}
have emerged as a powerful approach for representing and querying massive and complex genomic databases\omiii{~\cite{Iqbal2012,marchet2021data,karasikov2020metagraph,karasikov2022lossless,karasikov2019sparse,danciu2021topology,fan2023fulgor,bradley2019ultrafast,hunt2024allthebacteria}}. 
Unlike traditional models, which represent genomic databases as a collection of disjoint, linear sequences, sequences in a graph are represented by graph walks~\cite{marchet2021data}. Genome graphs provide two fundamental benefits, making them indispensable, particularly in modern, population-scale genomics~\cite{danko2021global,karasikov2020metagraph,siren2021pangenomics,Sherman2020,taylor2024beyond}.
First, the graph topology, along with its associated metadata\footnote{Metadata can include the original species or samples of a graph node's sequence~\cite{Iqbal2012,marchet2021data}, associations with patient outcomes~\cite{karasikov2020metagraph}, and more.} offer vastly \emph{greater expressive power}~\cite{eizenga2020pangenome,Sherman2020,taylor2024beyond,Liao2023}. A graph naturally encodes the evolutionary history and diversity of the organisms in a database, revealing their shared and distinct sequences~\cite{eizenga2020pangenome,Sherman2020,Armstrong2020cactus,bradley2019ultrafast}. This leads to reduced bias and improved analysis accuracy. 
For example, genome graphs improve disease diagnosis by capturing population-specific variants often missed by traditional, linear sequences~\cite{Groza2024,Sherman2020}.
Second, genome graphs leverage the inherent redundancy of genomic data to \emph{avoid redundant computation} on shared sequences. This is because shared genomic regions across database entries are stored only once and do not need to be queried separately. 
This is essential in settings like population-scale pathogen surveillance, where public health systems must rapidly match new \omii{pathogen} genomes against large national databases~\cite{danko2021global,karasikov2020metagraph,alipanahi2020metagenome,hunt2024allthebacteria,Vieira2024,gangwar2025wepp}.

\head{Data Movement Overhead} 
\asp{Due to the significance of graph-based genome analysis and its computational challenges, many works (e.g.,\omii{~\cite{\citegraphhwoptimization}}) propose techniques to alleviate these challenges by mitigating computational and main memory bottlenecks. However, to our knowledge, none of them mitigates the I/O cost of reading large amounts of graph data from the storage system to main memory and computation units.
Our motivational analysis (\sect{\ref{sec:motivation}}) shows that the end-to-end performance of analysis with large-scale genome graphs suffers significantly from the data movement overhead of accessing large amounts of low-reuse data (exacerbated by reduced access locality due to the graph structure).}

The I/O overhead of moving large amounts of low-reuse genome graph data is hard to avoid. One might think it is possible to eliminate this overhead by \inum{i}~shrinking graphs through sampling  (e.g.,~\cite{kim2016centrifuge,wood2019improved,muller2017metacache,song2024centrifuger,Dilthey2019,Fan2021}) or \inum{ii}~maintaining all required graphs completely and always in main memory. However, neither solution is suitable. The first approach necessarily reduces accuracy~\cite{berger2023navigating}, making it inapplicable for many 
use cases (e.g.,~\cite{milanese2019microbial,salzberg2016next,gihawi2023major,pockrandt2022metagenomic,berger2023navigating,Ackelsberg2015,Nasko2018,meyer2021critical,alser2020technology,alser2022molecules}).
The second approach is not suitable
since \asp{\inum{i}~genome graph databases, which are already large (tens to hundreds of terabytes~\cite{karasikov2020metagraph,metagraphaws}), grow in size rapidly due to increasing demands in biomedicine~\cite{clark2019diagnosis,farnaes2018rapid,sweeney2021rapid,ginsburg2009genomic,chin2011cancer,Ashley2016,bloom2021massively,gilchrist2015whole} and the exponentially decreasing cost of sequencing~\cite{nhgri}, and \inum{ii}~regardless of individual graph sizes, different analyses need to access different graphs.}
Therefore, it is energy-inefficient, unsustainable, costly, and unscalable to maintain \emph{all} data required for \emph{all possible} analyses in DRAM at all times.

\noindent\textbf{Storage-centric computing (SCC)}
refers to processing data inside the storage device, either on the SSD\footnote{In this work, we focus on the predominant NAND flash-based SSDs~\cite{cai2017error,cai-insidessd-2018}. We expect that our designs and insights would benefit storage systems with other emerging technologies 
\omii{(e.g.,~\cite{meena2014overview,lee2009architecting,akinaga2010resistive,tehrani1999progress})}
as well.} controller (in-storage processing, i.e., ISP) or on the flash dies (in-flash processing, i.e., IFP). SCC can be a fundamental solution for alleviating data movement overheads in graph-based genome analysis by processing data where it originally resides. This is due to three reasons.
First, SCC reduces unnecessary data movement between the storage system and compute units by handling large amounts of low-reuse data within the storage system.
Second, SCC alleviates the overall execution burden of moving and analyzing low-reuse data from the rest of the system (e.g., processing units and main memory), \omii{such that these components can be used for other purposes or be turned off to save energy.}
Third, SCC exploits the large internal bandwidth of the SSD,
which is particularly advantageous when computation is performed within the flash dies.

\head{Limitations of Prior Work} To our knowledge, no prior work alleviates the I/O overheads of graph-based genome analysis. Some works (e.g.,~\cite{Cali2022SeGraM,Zhang2024Harp,Zeng2024asgdp,Shen2024128parallel,Li2024,Mandal2020,Varma2013,Awan2021,Feng2021,Zhang2025,kim2025nmp,Huang2023meg2,Angizi2020Panda,Qiu2017}) accelerate graph-based genome analysis by alleviating its computation or main memory bottlenecks. 
However, they do not alleviate I/O overheads, whose impact on end-to-end performance gets even larger as other overheads are alleviated.
\asp{Several works propose SCC to} alleviate the I/O overheads of \inum{i} \emph{non-genomic} graph analysis (e.g.,\omii{~\cite{jun2018grafboost,matam2019graphssd,Wang2024ndsearch,lee2022smartsage,Niu2024flashgnn,Lee2024presto,Khadirsharbiyani2024smartgraph,Zhang2025taijigraph,An2023baraddur,Kang2024sting,xu2019vstore,li2021glist,kang2023near,wang2024beacongnn,kwon2025graphaccel,niu2022flashwalker,xu2026proxima}}) or \inum{ii}~genome analysis on conventional, \emph{linear sequences} \asp{(e.g.,\omii{~\cite{mansouri2022genstore,abakus23taco,megis,jun2016storage,kim2025nmp,soysal2025mars,Zheng2025ispgenome,tsai2025accelerating}})}. However, as detailed in \sect{\ref{sec:related}}, they are not directly applicable to the unique structures, access patterns, and 
operations on genome graphs.

\head{Challenges} Despite the benefits of SCC, no existing tool for graph-based genome analysis can be efficiently implemented in the SSD due to the limitations of hardware resources in modern SSDs. This is because graph-based genome analyses involve extensive random, irregular accesses to genome graphs, causing costly contention in internal SSD components (e.g., channels and NAND flash dies~\cite{nadig2023venice,kim2022networked,tavakkol2018flin}). This hinders leveraging the SSD's internal bandwidth, thereby diminishing the benefits of SCC.

\noindent\textbf{Our goal} in this work is to improve the performance and energy efficiency of graph-based genome analysis by reducing the large data movement overhead from the storage system in a low-cost way. To this end, we propose \proposal, the \emph{first} system for analysis with large-scale \underline{\textbf{g}}enome g\underline{\textbf{ra}}phs \underline{\textbf{in}} \underline{\textbf{s}}torage. 
\asp{\proposal is versatile and flexible as it supports \omii{major} operations on genome graphs (e.g.,  k-mer set lookup and read mapping) and can be integrated into various analysis pipelines.}
\asp{Through our detailed examination of typical graph-based genome analysis pipelines, we perform storage-aware algorithm-architecture co-design to \inum{i}~make the graph-based genome analysis pipelines more storage-friendly and \inum{ii}~further improve performance, energy-efficiency, and cost-effectiveness via SCC.}

\head{Key Mechanism}{} \proposal{}'s\omiiiq{\textbf{I kept the "s" as we did the same in MegIS and maybe we keep it consistent here (thus in thesis) as well? I can remove if too ugly}} \omii{algorithm-architecture} co-design is based on three key aspects.
\asp{First, we design an efficient batching technique and pipelined execution flow specialized to the properties of genome graphs and queries to reduce random accesses to the genome graph structure}. 
\asp{Second, we devise lightweight hardware units in the flash dies to process the needed parts of each page's graph data. This simple IFP design avoids unnecessarily transferring low-reuse or unused parts of flash pages outside the flash die, preventing SSD channel and external I/O bandwidth waste}.
\asp{Third, to fully exploit die-level parallelism in IFP, we design an effective, yet lightweight, scheduling technique, enabled by repurposing existing SSD structures. 
\proposal{}'s data layout drastically reduces the required in-DRAM mapping metadata, allowing us to use the freed-up internal DRAM space to maintain small per-die scheduling tables. We devise small ISP units on the SSD controller to interface with these in-DRAM tables and schedule IFP operations at low cost.}
\asp{To efficiently realize this SCC design, while guaranteeing correctness and integrity, 
we apply simple changes to the flash translation layer (FTL) and adopt a lightweight error correction scheme~\cite{lee2025aif}}.

\head{Key Results} 
On systems with two state-of-the-art SSD configurations, 
we compare \proposal against state-of-the-art software and hardware tools for graph-based genome analysis: \inum{i}~Fulgor: a node-centric tool~\cite{fan2023fulgor}, \inum{ii}~the MetaGraph framework: an edge-centric tool~\cite{karasikov2020metagraph,karasikov2022lossless,danciu2021topology,karasikov2019sparse}, and \inum{iii}~an \emph{idealized} hardware-accelerated tool without main memory overheads (IdealAccMem) to query the graph.
We show that when performing k-mer set lookups and alignment-free read mapping (two core operations in large-scale genome graph analyses), 
\proposal provides 6.8$\times$ (11.2$\times$), 10.7$\times$ (21.5$\times$), and 5.6$\times$ (9.8$\times$) average speedup (energy reduction) over Fulgor, MetaGraph, and IdealAccMem, respectively. 
\asp{\proposal{}'s workflow also seamlessly supports alignment-based mapping,
where the candidate graph regions identified by \proposal{} are fed to an alignment engine.} We integrate \proposal, Fulgor, and MetaGraph with SeGraM, a state-of-the-art sequence-to-graph alignment accelerator~\cite{Cali2022SeGraM}.
SeGraM+\proposal performs 6.2$\times$ and 9.0$\times$ better than SeGraM+Fulgor and SeGraM+MetaGraph.

This work makes the following \textbf{key contributions}:

\squishlist
    \item We demonstrate the impact of I/O overheads on the end-to-end performance of graph-based genome analysis.
    \item We introduce \proposal, the first system for analysis with large-scale genome graphs in storage. 
    \item \asp{We perform storage-aware algorithm-architecture co-design to make graph-based genome analysis  more storage-friendly, and further improve performance, efficiency, and cost-effectiveness via storage-centric computing (SCC).} 
    \item We show that \proposal improves performance and energy efficiency over the state-of-the-art graph-based genome analysis tools \omii{and accelerators}, without relying on costly resources throughout the system, thereby enabling wider adoption of graph-based genome analysis.  
\squishend

\section{Background}
\label{sec:background}

\subsection{Genome Sequence Analysis}

Given a genomic sample, a typical workflow consists of three key steps: \inum{i} sequencing, \inum{ii} basecalling, and \inum{iii} analysis. The first step, \emph{sequencing}, reads and converts the sample's DNA\footnote{Even if an organism's genome is RNA-based, it is typically converted to DNA before being sequenced~\cite{Houldcroft2017,Jansz2024viral}.} to digital signals. Modern sequencers cannot read long DNA molecules as complete sequences. Instead, they produce many shorter, overlapping sequences called \emph{reads}.
The second step, \emph{basecalling}, converts each raw read signal into a string, encoded using 
\texttt{A}, \texttt{C}, \texttt{G}, \texttt{T}~\cite{cock2009sanger}.
The third step is \emph{genome analysis} on a set of reads (i.e., a \emph{read set}).
\asp{Common genome analysis workflows then involve querying reads from 
samples against a single \emph{reference} genome or large-scale databases (e.g., containing reference genomes from many species or previously sequenced samples) to determine species or genomic features present in the sample~\cite{karasikov2020metagraph,alanko2023themisto,Muggli2017SuccinctGraphs,fan2023fulgor,piro2020ganon,mangul2016reference} and/or to find their differences from genomes in the database~\cite{aganezov2022complete,Groza2024}.}

In this work, we focus on the analysis performance for two reasons. 
First, sequencing and basecalling are typically performed once per read set, whereas analysis of the same read set can be performed \emph{many times} or at \emph{different times}.
For example, read sets from prior disease-association studies are commonly reanalyzed as human or microbial genome databases improve~\cite{rhie2023complete,miga2020telomere,taylor2024beyond,aganezov2022complete,chen2024improved} or expand~\cite{taylor2024beyond,hunt2024allthebacteria}, or when new personalized databases can be tailored to the study cohort~\cite{Siren2024,Sherman2020}.
Second, modern sequencing throughput far outpaces Moore's law~\cite{nhgri,stephens2015big,berger2023navigating}, which poses significant computational challenges for analysis on conventional systems. This is why genome analysis tasks have been a key acceleration target in a large body of works (e.g.,~\cite{\citehwoptimization,\citemghwoptimization}).

\subsection{Graph-Based Genome Analysis}
\label{sec:background-graphs}

\noindent\textbf{Genome Graphs}
have emerged as a powerful approach for representing and querying massive and complex genomic databases\omiii{~\cite{Cali2022SeGraM,marchet2021data,karasikov2020metagraph,fan2023fulgor,karasikov2022lossless,karasikov2019sparse,danciu2021topology,bradley2019ultrafast,Muggli2017SuccinctGraphs,Iqbal2012,turner2018integrating,hunt2024allthebacteria}}. Unlike traditional models, which represent genomic databases as a collection of disjoint, linear sequences, sequences in a graph are represented by graph walks\omiii{~\cite{marchet2021data,Iqbal2012}}. Genome graphs collapse subsequences shared across different database entries into single nodes and indicate adjacent subsequences with edges, compacting sequence sets (by up to thousand-fold~\cite{karasikov2020metagraph}).
A node's metadata then indicates which entries contain the represented sequence.

Genome graphs provide two fundamental benefits, which make them indispensable, particularly in modern, population-scale genomics~\cite{danko2021global,karasikov2020metagraph,siren2021pangenomics,Sherman2020,taylor2024beyond}.
First, the graph topology, along with its metadata, offer \emph{\textbf{\omiii{vast} expressive power}}~\cite{eizenga2020pangenome,taylor2024beyond,Liao2023,Sherman2020,Mustafa2024MLA}. A graph naturally encodes the evolutionary history and diversity of a cohort of organisms, revealing their shared and distinct sequences~\cite{eizenga2020pangenome,Armstrong2020cactus,bradley2019ultrafast,Sherman2020}. 
This leads to reduced bias and improved accuracy in analyses. For example, genome graphs improve disease diagnosis by capturing population-specific variants often missed by analyses on traditional, linear sequences~\cite{Groza2024,Sherman2020,jain2019pasgal,Negi2025}. Second, genome graphs leverage the inherent redundancy of genomic data to \emph{\textbf{avoid redundant computation}}.  This is because shared genomic regions across database entries are stored only once and do not need to be queried separately. This is essential in settings such as population-scale pathogen surveillance, where public health systems must rapidly match new pathogen genomes against large national databases~\cite{danko2021global,karasikov2020metagraph,alipanahi2020metagenome,hunt2024allthebacteria,Vieira2024,gangwar2025wepp}, enabling timely detection of local outbreaks and limiting broader spread.

\begin{figure}[t]
         \centering
\includegraphics[width=0.93\columnwidth]{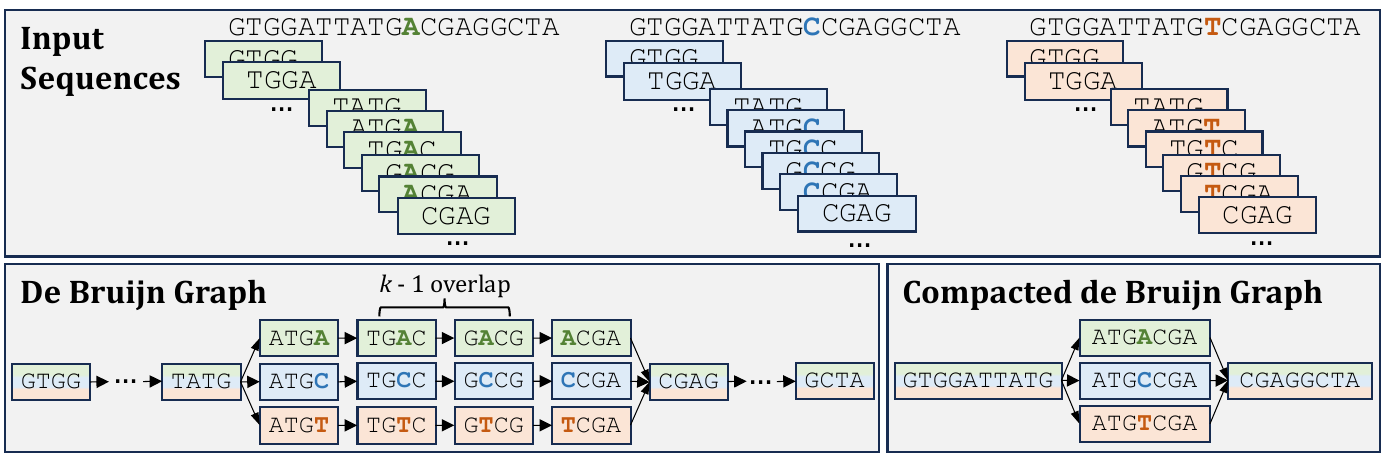}        
  \caption{An example de Bruijn graph and its compacted form.}
\label{fig:dbg}
\end{figure}

\head{De Bruijn Graphs} To efficiently represent large-scale genome graphs, \emph{de Bruijn graphs} (DBGs), which are a type of overlap graph\omiii{~\cite{Cali2022SeGraM,marchet2021data,karasikov2020metagraph,fan2023fulgor,karasikov2022lossless,karasikov2019sparse,danciu2021topology,bradley2019ultrafast,Muggli2017SuccinctGraphs,Iqbal2012,turner2018integrating,pevzner_eulerian_2001,mustafa2022algorithms}}, have been adopted by many recent works~\cite{Muggli2017SuccinctGraphs,fan2023fulgor,alanko2023themisto,cracco2023extremely,karasikov2022lossless,danciu2021topology,karasikov2019sparse,karasikov2020metagraph,bradley2019ultrafast,Iqbal2012,turner2018integrating}.
While other genome graph classes exist~\cite{Baaijens2022,paten2017genome,Armstrong2020cactus},
DBGs scale much better
 \omiii{with both the entries' size and genetic diversity of genomic} databases
\cite{karasikov2020metagraph,bvrinda2023efficient,Chikhi2013,karasikov2019sparse}, which is why they are preferred in these contexts. \fig{\ref{fig:dbg}} shows an example of a DBG, where each node is a unique \emph{k-mer} (i.e., a subsequence of length $k$) that is observed in the input, and there is a directed edge from node $u$ to node $v$ if the suffix $(k-1)$-mer of $u$ is equal to the prefix $(k-1)$-mer of $v$. 
Each node is then assigned a \emph{color}, representing its metadata. 
A common strategy to reduce a DBG's size is to instead store its corresponding \emph{compacted DBG}, where all $k$-mers from a maximal non-branching path are merged into a single node. 
There are two common ways to represent (compact) DBGs~\cite{alanko2023small}.
A \emph{node-centric} DBG of order $k$ (e.g.,~\cite{fan2023fulgor,pibiri2022sparse,alanko2023themisto}) stores all $k$-mers observed in the entries and defines its edges implicitly (\fig{\ref{fig:dbg}}). An \emph{edge-centric} DBG of order $k$ (e.g.,~\cite{Bowe2012SuccinctGraphs,Muggli2017SuccinctGraphs,li2015megahit,karasikov2020metagraph,karasikov2022lossless,danciu2021topology,karasikov2019sparse}) stores observed $(k-1)$-mers as nodes and only stores the edges that represent observed $k$-mers. In the rest of this paper, we refer to compact DBG as DBG for short.

\begin{figure}[h]
         \centering
         \includegraphics[width=\columnwidth]{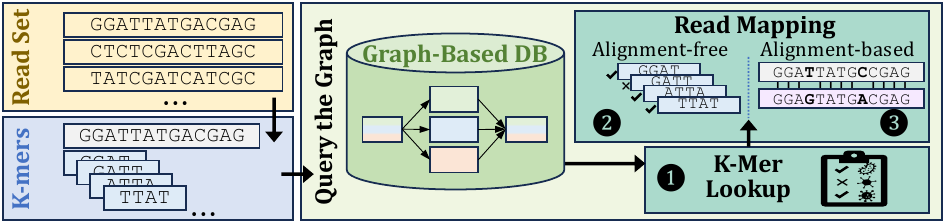}
         \caption{\asp{High-level overview of graph-based genome analysis.}}
         \label{fig:analysis-bg}
\end{figure}

\noindent\textbf{{\asp{Graph-based genome analysis}}} 
involves 
querying reads against the graph to determine what species or genomic features are present in the sample~\cite{karasikov2020metagraph,alanko2023themisto,Muggli2017SuccinctGraphs,fan2023fulgor,piro2020ganon,mangul2016reference} and/or to find their differences from genomes in the database~\cite{aganezov2022complete,Groza2024}.
As shown in \fig{\ref{fig:analysis-bg}}, queries fall into two categories: \circled{1} \emph{k-mer set lookups} (where k-mers in a read set are matched against the graph to determine the species present in a sample), and \emph{read mapping} (where each read is analyzed individually). Read mapping can be done by either \circled{2} alignment-free or \circled{3} alignment-based approaches. Alignment-free approaches (more common for large-scale studies~\cite{karasikov2020metagraph,karasikov2022lossless}) match all k-mers of a read to the graph, retrieve the metadata associated with the graph nodes to which the k-mers match, and classify the read using this metadata. Alignment-based approaches introduce an additional, computationally expensive approximate string matching step, where candidate regions identified by k-mer matches are refined via dynamic programming to determine precise differences between the read and the graph. 

\subsection{Unique Characteristics of Genome Graphs}
\label{sec:graph-uniqueness}

The type of data encoded in genome graphs and the underlying goals of querying genome graphs exhibit unique features not encountered in conventional, non-genome graphs (e.g., social networks, web graphs, road networks). General-purpose graph data structures and methods do not account for these features, leading to significant inefficiencies in storage and computation requirements and missed opportunities. This has led to significant efforts in both software (e.g.,~\cite{rautiainen2020graphaligner,kim2019hisat2,gao2020abpoa,jain2019pasgal,siren2021pangenomics,Rautiainen2019,Chandra2023,Ivanov2022,Ma2023,Darby2020vargas,Hwang2025MEMO,Romain2023svjedi,Li2020minigraph}) and hardware (e.g.,~\cite{Cali2022SeGraM,Zhang2024Harp,Zeng2024asgdp,Shen2024128parallel,Li2024,Mandal2020,Varma2013,Awan2021,Feng2021,Zhang2025,kim2025nmp,Huang2023meg2,Angizi2020Panda,Qiu2017}) to develop specialized techniques for genome graphs.

The unique semantic characteristics in genome graphs manifest in three key aspects regarding the graph structure,  storage (e.g., sparsity and degree distribution), and access patterns. First, in a traditional graph with $N$ nodes, there are $\mathcal O(N)$ to $\mathcal O(N^2)$ edges that must all be stored explicitly (e.g., adjacency lists\omiii{~\cite{cormen2009introduction,hopcroft1973algorithm}}, \omiii{compressed sparse row formats~\cite{ligra,merrill2012scalable}}). 
Genome graphs, most commonly represented as DBGs, are structurally constrained in that they either store nodes or edges (as detailed in \sect{\ref{sec:background-graphs}}) and the other is obtained implicitly. This is a feature necessitated by the massive scale of genome graphs and enabled by the inherent overlap structure of the sequences they encode.

Second, traditional graphs typically exhibit power-law distributions, \omiii{in which a small number of nodes have very high degree and many graph-processing optimizations (e.g.,~\cite{chi2022accelerating,Dadu2021polygraph,yao2022scalagraph}) exploit this skew}. However, DBG node outdegrees and indegrees \omiii{are} at most four (determined by the fixed alphabet size, \texttt{A}, \texttt{C}, \texttt{G}, \texttt{T}, in the underlying data) regardless of the overall graph size. Therefore, optimizations that exploit skewed degree distributions, a common strategy in conventional graph processing systems, are \omiii{not well-suited} to genome graphs, necessitating different optimization approaches.

Third, due to the differences in underlying graph structures, access pattern optimization opportunities vary between genome graphs and non-genome graphs. The compressed data structures used in large-scale genome graphs enforce specific node orderings and, thus, \omiii{restrict} node reordering, a technique commonly used in non-genome graphs to improve access patterns~\cite{Bowe2012SuccinctGraphs,pibiri2022sparse}. However, the fact that genome graphs encode biological sequences, and that queries are themselves biological sequences, opens new avenues for improving access pattern locality that are unavailable in conventional graph workloads. For example, k-mers from different query reads that share common substrings map to nearby regions in the graph's index structures. In contrast, most conventional non-genome graph queries do not exhibit this kind of structural coupling between the queries and the graph, limiting opportunities for analogous locality-aware optimizations.

\subsection{SSD Organization}
\label{sec:background-ssd}

\begin{figure}[ht]
         \centering
         \includegraphics[width=0.77\columnwidth]{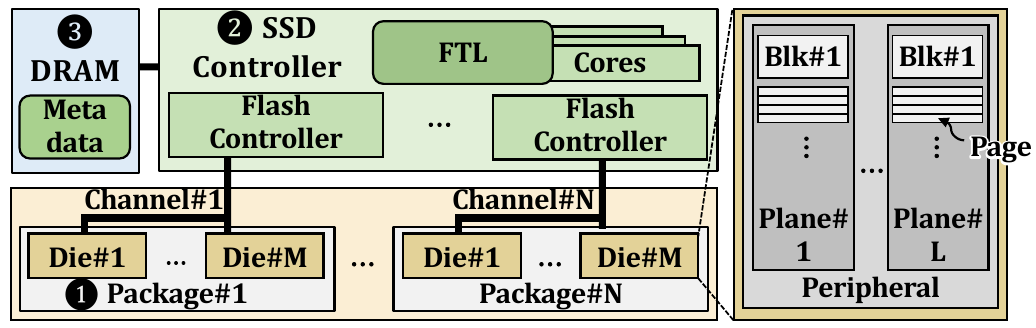}
         \caption{Organizational overview of a modern SSD.}
         \label{fig:ssd}
\end{figure}

\fig{\ref{fig:ssd}} depicts the organization of a modern SSD.

\head{\circled{1}~NAND Flash Memory}
A NAND package consists of multiple \emph{dies} or \emph{chips} that share the NAND package's I/O. 
One or multiple packages share a command/data bus or \emph{channel} to communicate with the SSD controller\omiii{~\cite{micheloni2010inside,cai-insidessd-2018,cai2018errorsarxiv,cai2017error,cai2015read,cai2017vulnerabilities,cai_data_2015,tavakkol2018mqsim,nadig2023venice}}.
Each channel can be used by only one die at a time,
 to communicate with the controller,
whereas dies can operate independently.
Each die has multiple (e.g., 2 or 4) \emph{planes}, each with thousands of \emph{blocks}, and
each block has hundreds to thousands of 4--16 KiB \emph{pages}.

Read/write operations take place at page granularity\omiii{;} erase operations at block granularity.
The planes in each die share the peripheral circuitry for access to pages.
Thus, planes in a die can operate concurrently when accessing pages (or blocks) at the same offset via \emph{multiplane} operations\omiii{~\cite{micheloni2010inside,cai-insidessd-2018,cho2024aero,mansouri2022genstore,park2022flash}}.

\head{\circled{2}~SSD Controller}
An SSD controller consists of two key components: the \emph{flash translation layer (FTL)} and the \emph{flash controllers.}  
The FTL, executed on multiple cores, 
is responsible for communication with the host, internal I/O scheduling, and various SSD management tasks. The \emph{per-channel hardware flash controllers} manage the request handling\omiii{~\cite{micheloni2010inside,cai-insidessd-2018,cai2018errorsarxiv,cai2017error,cai2015read,cai2017vulnerabilities,cai_data_2015,tavakkol2018mqsim,nadig2023venice,cai2013program}} and error correction for the NAND flash chips\omiii{~\cite{dong2010use,zhao2013ldpc,bose1960class,wang2014enhanced,cai2017vulnerabilities, luo2018improving,luo-hpca-2018,cai2015read,cai2013error,cai2017error, ha2015integrated,luo2015warm,cai_data_2015,cai_error_2012,cai2018errorsarxiv,cai2013program,cai2012flash,cai-insidessd-2018}}.

\head{\circled{3}~DRAM} Modern SSDs employ low-power DRAM to store metadata crucial for SSD management, e.g., logical-to-physical (i.e., L2P) mappings. %
L2P mappings are typically maintained at a granularity of 4KiB to enhance random access performance. 
The required capacity for L2P mappings is about 0.1\% of the SSD's capacity, considering a 32-bit architecture, with 4B of metadata for every 4KiB of data. 
This translates to a 4-GB LPDDR4 DRAM for a 4-TB SSD~\cite{samsung860pro}.

\section{Motivational Analysis}
\label{sec:motivation}

We conduct experimental analyses to assess the impact of I/O overheads on graph-based genome analysis performance.

\subsection{Data Movement Overheads}
\label{sec:motivation-dm}

\head{Tools and Datasets} We use two state-of-the-art tools for large genome graph analysis: \inum{i}~Fulgor~\cite{fan2023fulgor}, a node-centric tool, and \inum{ii}~the MetaGraph framework~\cite{karasikov2020metagraph,karasikov2022lossless,danciu2021topology,karasikov2019sparse}, an edge-centric tool. For each, we use the best-performing thread count and color encoding scheme. We evaluate k-mer set lookup, a fundamental task on genome graphs (\sect{\ref{sec:background}}). 
Graphs are built from a pilot subsample of the global MetaSUB Consortium~\cite{danko2021global} dataset.
The resulting graph (with colors) is 659\,GB for Fulgor and 822\,GB for MetaGraph.\omiiiq{\tiny You asked whether we can evaluate even larger graphs. Constructing individual larger graphs is very challenging (needs much larger DRAM than our max 1.5TB DRAM nodes, and takes a very long time). But we can analyze databases of several graphs, which is also a common practice in reality. We already evaluate databases of multiple graphs in \sect{\ref{sec:evaluation}}, but we can scale the overall size even further.\\}  
We evaluate queries with varying read counts, ranging from 100K reads\footnote{Note that smaller queries are also critical, as some applications (e.g., searching for antimicrobial resistance~\cite{danko2021global,karasikov2020metagraph,alipanahi2020metagenome,hunt2024allthebacteria} or for a single gene of interest~\cite{edgar2022petabase}) require searching a single or a few genes within a large graph.} to 10M reads (35 MB to 3.5 GB).
\sect{\ref{sec:methodology}} provides further methodology details.

\head{System Configurations}
Our experiments run on a high-end server with an AMD EPYC 7742 CPU~\cite{amdepyc} and 1.5 TB of DDR4 DRAM~\cite{ddr4sheet}. DRAM capacity \emph{exceeds} the total size of all data accessed during the analysis. 
This ensures that \irevminor{we} capture the intrinsic I/O overhead of transferring large, low-reuse data from the storage system to main memory, without being constrained by memory capacity. We analyze
I/O overheads 
using two state-of-the-art SSDs: \inum{i}~\ssdm, with a PCIe Gen4 interface~\cite{samsungPM1735} and \inum{ii}~\ssdh, with a PCIe Gen5 interface~\cite{samsung9100PRO}.

\begin{figure}[b]
         \centering
         \includegraphics[width=0.9\columnwidth]{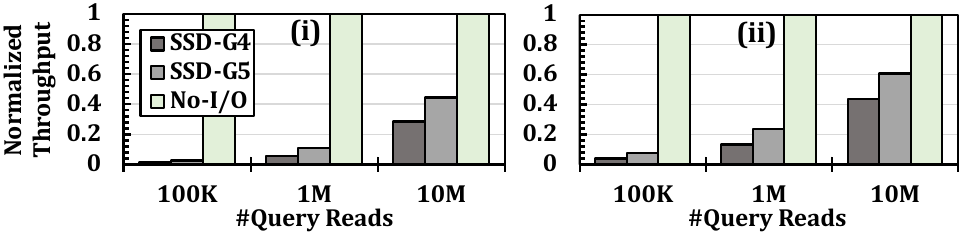}
         \caption{Normalized throughput of \inum{i}~Fulgor and \inum{ii}~MetaGraph under different storage configurations and query sizes.}
         \label{fig:motivation1}
\end{figure}

\head{Observations}
\fig{\ref{fig:motivation1}} shows throughput (\#queries/sec) of the tools, normalized to that of a hypothetical configuration with zero performance overhead due to storage I/O (\dram).  
We observe that in all cases, I/O leads to large overheads, even with the state-of-the-art SSDs. Compared to \ssdm (\ssdh), \dram leads to 16.7$\times$ (9.3$\times$) and 7.5$\times$ (4.5$\times$) better average performance in Fulgor and MetaGraph, respectively.

\begin{figure}[h]
         \centering
         \includegraphics[width=0.9\columnwidth]{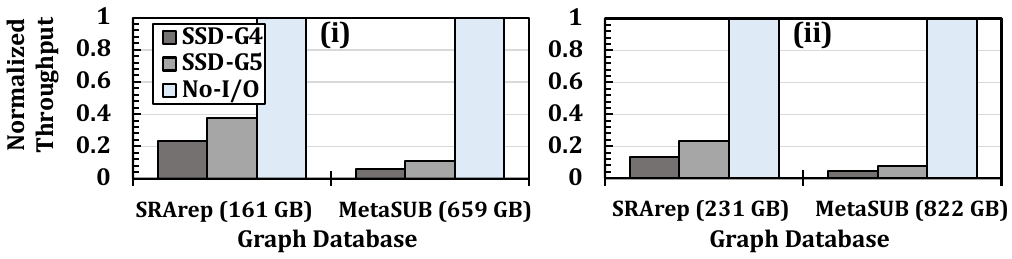}
         \caption{Normalized throughput of \inum{i}~Fulgor and \inum{ii}~MetaGraph under different storage configurations and graph sizes.}
         \label{fig:motivation2}
\end{figure}

\fig{\ref{fig:motivation2}} shows the throughput of the tools normalized to \dram, across different database sizes. The larger graph (G\_MetaSUB) is the one discussed earlier. The smaller graph (G\_SRArep) is generated from a representative subset of the Sequencing Read Archive's~\cite{katz2021sra} public portion. 
The resulting graph (with colors) is 161 GB for Fulgor and 231 GB for MetaGraph.  We observe that, in all cases, I/O overhead is substantial and increases with database size. For example, the speedup of \dram over \ssdh increases from 2.7$\times$ to 9.2$\times$ \omiii{(as the graph database size grows from 161 GB to 659 GB)} in Fulgor and from 4.3$\times$ to 13.4$\times$ \omiii{(as the graph database size grows from 231 GB to 822 GB)} in MetaGraph.

Based on these observations, we conclude that storage I/O leads to large overheads in graph-based genome analysis, an overhead that is expected to exacerbate in the future. This I/O overhead is due to moving large amounts of \emph{low-reuse} data from the storage system all the way to main memory, caches, and computational units. Due to its low reuse, caching this data in DRAM during analysis (even if DRAM is larger than the data size, as evaluated in our analyses) does not significantly amortize this overhead. While several works (e.g.,~\cite{\citegraphacc}) accelerate analysis with genome graphs, to our knowledge, they do not address storage I/O overheads. Although mitigating other bottlenecks (e.g., main memory or computation) leads to large benefits, \omiii{doing so does} not alleviate I/O overheads, whose impact on end-to-end performance becomes even larger as other bottlenecks are alleviated. 

This I/O overhead, due to moving large amounts of low-reuse data, is hard to avoid. One might think it is possible to avoid it by \inum{i}~producing smaller databases through sampling  (e.g.,~\cite{kim2016centrifuge,wood2019improved,muller2017metacache,song2024centrifuger,Dilthey2019,Fan2021}) or \inum{ii}~maintaining all required graph data completely and always resident in main memory. However, neither of these is suitable. The first approach necessarily reduces accuracy~\cite{berger2023navigating}, rendering it inapplicable for many use cases (e.g.,~\cite{\citegraphacc}). The second approach is energy-inefficient, unsustainable, costly, and unscalable for two reasons. First, the sizes of graph-based databases  (which are already large, i.e., tens to hundreds of TBs in some recent examples~\cite{karasikov2020metagraph,metagraphaws}) have been growing \omiii{rapidly}~\cite{katz2021sra,enastats,nhgri,stephens2015big}. Second, independent of the sizes of individual graphs, different analyses need different graphs, constructed from different sets of genomes and/or with varying parameters. For example, clinical studies may require graphs with patients' genomes kept separately for privacy~\cite{Berger2019}, while metagenomic studies and wastewater monitoring require highly diverse genome sets~\cite{gihawi2023major,song2024centrifuger,parkins2024wastewater,Blanco-Miguez2023}. Therefore, keeping all data for all possible analyses in DRAM at all times is prohibitively inefficient and ultimately unsustainable.

Several works (e.g.,~\cite{mansouri2022genstore,abakus23taco,megis,soysal2025mars}) have already demonstrated significant I/O overheads in non-graph-based genome analysis. While genome graphs enable a more compact representation of population-scale databases (thereby making it feasible to analyze databases that would be prohibitively large under traditional, linear representations), they still suffer from large I/O overheads. This is because the massive and continually growing scale of these databases still leads to very large graphs with low data reuse, and the graph topology introduces irregular, dependent access patterns, further reducing locality.

\vspace{-0.5em}
\subsection{\asp{Alleviating Data Movement Overheads}}
\label{sec:motivation-ch-goals}

\subsubsection{Storage-Centric Computing (SCC)}
Processing data in the storage device can be a fundamental approach for reducing the I/O overheads for three reasons. First, SCC eliminates unnecessary data movement of low-reuse data across the system. Second, SCC reduces the computational burden imposed by low-reuse data on the rest of the system (e.g., main memory and computational units), \omii{such that these components can be used for other purposes or be turned off to save energy, \omiii{or can be made simpler or less costly~\cite{megis,boroumand_google_2021}}.} Third, as highlighted by many prior works (e.g.,\cite{li2023ecssd,mailthody2019deepstore,kang2021mithrilog,koo2017summarizer,mansouri2022genstore,wang2024beacongnn,jang2024smart,Kim2023optimstore,li2021glist}), SCC can leverage the high internal bandwidth of SSDs. In modern SSDs, the internal bandwidth often surpasses the external bandwidth. For example, a state-of-the-art SSD controller\cite{flashtecnvme5016} delivers 14~GB/s of external bandwidth and up to 57.6~GB/s of internal bandwidth (achieved via 16~channels, each with a peak of 3.6~GB/s). Over-provisioning internal bandwidth in modern SSDs is important to protect user-perceived external I/O performance, mitigating the effects of \inum{i} channel contention~\cite{nadig2023venice,kim2023decoupled,kim2022networked,tavakkol2014design} and \inum{ii} the SSD's internal data migration for management tasks~\cite{tavakkol2018flin,kim2020evanesco,cai2017error,park-dac-2019} and refresh~\cite{cai2013error,luo2018improving}. The effective internal bandwidth increases even further when processing directly in the NAND flash dies~\cite{chun2022pif,park2022flash,lee2025aif}. This is because in-flash processing exploits the aggregate bandwidth across all dies, a bandwidth that grows with the number of dies and significantly exceeds both \inum{i}~the bandwidth available to processing units on the SSD controller and \inum{ii}~the external SSD bandwidth.

\subsubsection{Challenges and Goal}
Designing a SCC system for graph-based genome analysis is challenging because none of the existing approaches can be directly implemented inside storage effectively due to modern SSDs' constrained hardware resources. Both node- and edge-centric representations of genome graphs require many random accesses during analysis. These random and irregular accesses cause costly contention in internal SSD components (e.g., channels and NAND flash chips~\cite{nadig2023venice,kim2022networked,tavakkol2018flin}), which hinder leveraging the SSD's internal bandwidth.
Therefore, directly adopting existing approaches inside storage leads to performance and energy overheads. 

\asp{Some SCC systems in other domains regularize accesses by performing sorting in either the SSD or the host. However, these approaches are not effective \omiii{for} graph-based genome analysis. Sorting the large number of accesses is impractical within the SSD due to resource constraints. Sorting on the host also fails to fully address access irregularity, as graph-based genome analysis involves dependent accesses. Coordinating sorting between the host and the SSD for each round of dependent accesses incurs significant data movement overhead.}

\textbf{Our goal} in this work is to improve the performance and efficiency of graph-based genome analysis by alleviating its data movement overheads from storage cost-effectively.

\section{\proposal}
\label{sec:mech}

We propose \emph{\proposal}, a versatile storage-centric system for graph-based genome analysis. \proposal supports \omii{major} operations in analysis with genome graphs (e.g., k-mer set lookup and read mapping), 
and for alignment-based workflows, it flexibly integrates with existing alignment accelerators.
\proposal is primarily designed as a system to accelerate analysis with genome graphs. \proposal augments the existing SSD controller, flash dies, and FTL, and when not in the analysis acceleration mode, the SSD remains available for other applications, like a conventional general-purpose SSD.

We address the challenges of designing a SCC system for graph-based genome analysis via storage-aware algorithm architecture co-design based on our detailed examination of typical analysis pipelines that operate on genome graphs. To this end, we \inum{i}~make these pipelines more storage-friendly and \inum{ii}~further improve their performance, energy-efficiency, and cost-effectiveness via ISP and IFP. 

\proposal{}'s \omii{algorithm-architecture} co-design comprises three key aspects. \asp{First, we design \emph{\textbf{efficient batching and reordering techniques and pipelined execution flow}} specialized to the properties of genome graphs to reduce the number of random accesses to the graph}. 
\asp{Second, we devise \emph{\textbf{lightweight IFP units}} to process the needed parts of each page's graph data in the flash die. This IFP design avoids transferring low-reuse or unused parts of flash pages outside the die, thereby preventing bandwidth waste}.
\asp{Third, to fully exploit die-level parallelism during IFP, we design an \emph{\textbf{effective yet lightweight SCC scheduling technique}}, enabled by repurposing existing SSD structures. 
\proposal{}'s data placement (\sect{\ref{sec:mech-ftl-ecc}}) drastically reduces the required in-DRAM mapping metadata, allowing us to use the freed-up internal DRAM space to maintain small per-die scheduling tables. 
We devise small ISP units on the SSD controller to interface with these tables and schedule IFP operations at low cost.}
\asp{To efficiently realize these aspects, while guaranteeing correctness and integrity, we apply simple changes to the FTL and adopt a lightweight ECC~\cite{lee2025aif}}.

\asp{Leveraging our optimizations, \proposal{}’s SCC steps require only simple operations and \omiii{small} buffer spaces, enabling execution on either \inum{i}~our lightweight, specialized IFP and ISP hardware units or \inum{ii}~general-purpose IFP (e.g.,\omiii{~\cite{chun2022pif,chen2024search,park2022flash,Chen2024aresflash}}) and ISP (e.g.,\omiii{~\cite{gu2016biscuit, kang2013enabling, wang2019project,acharya1998active,keeton1998case,riedel1998active,riedel2001active,merrikh2017high,tiwari2013active,tiwari2012reducing,boboila2012active,bae2013intelligent,torabzadehkashi2018compstor,kang2021iceclave,zou2022assasin,nadig2026conduit}}) units. Efficient genome graph analysis on both stems from \proposal{}’s specialized batching, scheduling, and data/computation flow coordination. Execution on special-purpose lightweight hardware leads to higher power efficiency (\sect{\ref{sec:evals-area}}), while execution on existing \omiii{general-purpose} cores on the SSD controller or general-purpose \omiii{IFP and} ISP  units facilitates near-term adoption. Ultimately, choosing between these \proposal{} configurations is a design decision.}

\begin{figure}[t]
         \centering         \includegraphics[width=\columnwidth]{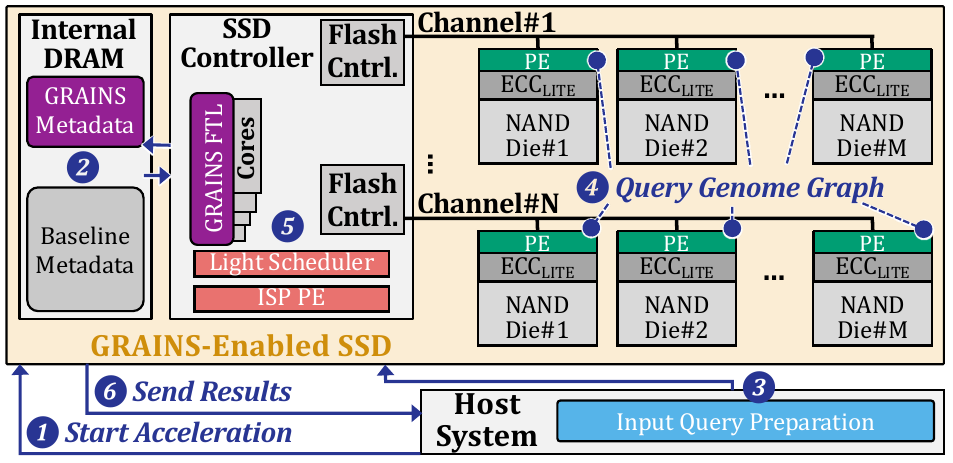}       
  \caption{Overview of \proposal.}
\label{fig:grains-overview}
\end{figure}

\fig{\ref{fig:grains-overview}} shows an overview of \proposal, with its lightweight IFP processing elements (PEs) on NAND flash dies and the lightweight ISP PE and scheduler on the SSD controller. \proposal FTL (\sect{\ref{sec:mech-ftl-ecc}}) orchestrates host communications and data flow across the SSD components. 
After receiving a request through \proposal's interface commands (\sect{\ref{sec:mech-interface}}) for accelerating graph-based genome analysis (\circled{1} in \fig{\ref{fig:grains-overview}}), \proposal prepares (\circled{2}) for SCC execution by flushing the conventional FTL metadata and loading the \proposal FTL metadata (\sect{\ref{sec:mech-ds-layout}}).
When execution starts, in the \emph{first step} (\sect{\ref{sec:mech-step1}}), \proposal prepares input queries in the host via efficient batching and pipelined execution, and transfers the query batches to the SSD (\circled{3}). 
In the \emph{second step} (\sect{\ref{sec:mech-step2}}), \proposal accesses graph data structures via its IFP (\circled{4}) and ISP (\circled{5}) units and sends the query results to the host (\circled{6}). These two steps are pipelined, with efficient coordination between resources to avoid writes during SCC. 
Through its efficient execution flow and by avoiding frequent management tasks (by not needing writes during SCC), \proposal effectively leverages the SSD's large internal bandwidth.

\section{Data Structures}
\label{sec:mech-ds-layout}

\asp{\proposal uses a node-centric DBG representation based on an associative dictionary to compactly represent graph sequences and their associated colors. Similar to prior large-scale genome graph analysis tools (e.g.,~\cite{campanelli2025fast,fan2023fulgor,Pibiri2024macdbg,campanelli2024where}), we use SSHash~\cite{pibiri2022sparse} as our dictionary since, by exploiting the statistical features of k-mers,
it achieves a substantially better space-time trade-off over prior 
sequence dictionaries.\footnote{\asp{Since our techniques rely on general DBG properties, they are also applicable to other k-mer dictionaries that may be used as DBG backbones\omiii{~\cite{karasikov2020metagraph,fan2023fulgor,alanko2023themisto,rautiainen2020graphaligner,campanelli2025fast,Pibiri2024macdbg,campanelli2024where,Muggli2017SuccinctGraphs,turner2018integrating}}.}}}

\begin{figure}[h]
         \centering
         \includegraphics[width=0.83\columnwidth]{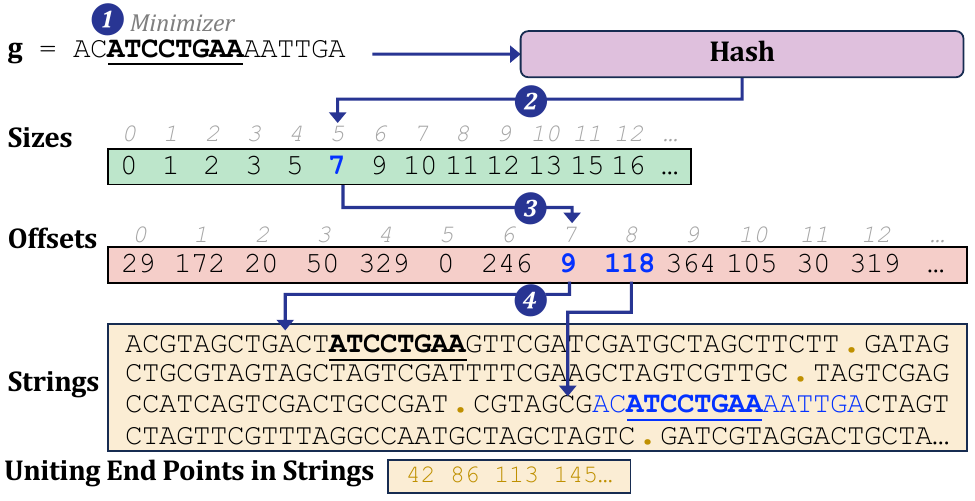}       
  \caption{Overview of the DBG structures and k-mer query.}
\label{fig:example-sshash}
\end{figure}

\head{Graph Sequences} We store DBG's \emph{unitigs} (i.e., maximal non-branching paths) as contiguous
strings in a predetermined order, so that a k-mer occurring in any unitig can be quickly located using a minimal perfect hash function~\cite{pibiri2021pthash} built for the set of k-mers.\footnote{Note that in DBGs, edges between unitigs are represented implicitly via $(k-1)$-mer overlaps (\sect{\ref{sec:background-graphs}}).}
\fig{\ref{fig:example-sshash}} shows an overview of the DBG structures. The unitigs are stored in \emph{\strings}, and are indexed via \emph{\offsets} and \emph{\sizes}. Given a query read $R$, k-mers from the read are extracted and looked up.
Consider k-mer $g$. First, its \emph{minimizer} (i.e., the first $m$-mer with the smallest hash value) is extracted (\circled{1} in \fig{\ref{fig:example-sshash}}). Second, with the minimizer's minimal perfect hash value ($h$), \sizes is indexed (\circled{2}). Third, with the \sizes values at locations $h$, \offsets is indexed (\circled{3}). $Sizes[h]$ from $Sizes[h+1]$ is then subtracted to find how many \offsets elements to read at 
$Sizes[h]$. Fourth, with the \offsets values, \strings is indexed (\circled{4}), \omiii{and} a window of length $k-m+1$~\cite{pibiri2022sparse} in \strings is checked to find the minimizer and check if the rest of the k-mer matches.

\begin{figure}[h]
         \centering
         \vspace{0.1em}
         \includegraphics[width=0.75\columnwidth]{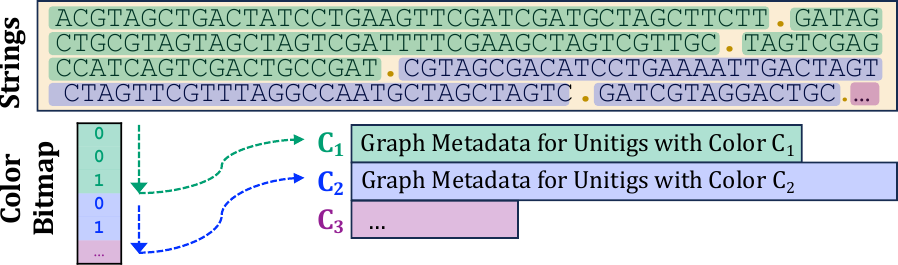}       
  \caption{Graph color \omiii{encoding used in \proposal}.}
  \vspace{-0.5em}
\label{fig:colors}
\end{figure}

\head{Graph Colors} For storing graph colors, we adopt an approach similar to~\cite{fan2023fulgor}. As shown in \fig{\ref{fig:colors}}, we sort the unitigs in \strings based on their colors (i.e., the ID pointing to the metadata for that graph unitig), and use a simple \emph{Color Bitmap} to mark the start of each new color. This encoding efficiently stores repeated colors at low \omiii{storage} overhead.    

\section{Input Query Processing}
\label{sec:mech-step1}

\asp{In this step, \proposal prepares reads for efficient graph queries in the next step (\sect{\ref{sec:mech-step2}}). We perform two new optimizations inspired by the \omiii{characteristics} of genome graphs and queries.}

\asp{\head{Cross-Read K-Mer Batching} Instead of querying each read in a read set individually, we analyze k-mers across different reads together. This way, we can batch k-mers to reduce the number of random accesses to the graph. To enable this, \proposal populates a lightweight data structure associating each k-mer with the reads it originates from. When graph queries return the colors of matched k-mers, \proposal assigns colors to each read. We perform this step in the host since extracting k-mers from reads incurs frequent writes, which would reduce the SSD's lifetime if performed inside the SSD.}

\asp{\head{Genome-Graph-Aware Query Reordering} 
We leverage a common property of minimal-perfect-hash–based k-mer dictionaries in DBGs: 
The metadata required for identifying the small per-minimizer lookup ranges (i.e., the \sizes array, as shown in \fig{\ref{fig:example-sshash}}) is much smaller than the graph’s \offsets and \strings.
Although the full graph is far too large and incurs large I/O overhead to access it in the host (as shown in \sect{\ref{sec:motivation}}), \sizes occupies only a small fraction of the total space (e.g., $<$4\% in our large-scale graphs). This enables a powerful opportunity: We perform k-mer lookups on \sizes in the host and use the resulting \offsets indices to sort and batch k-mers before sending them to the SSD, thereby reducing the number of accesses and improving the access patterns to \offsets.\omiiiq{You asked "also Strings?"\\No, Strings needs other techniques, as discussed in the next sections.} To execute this step efficiently, as detailed below, we ensure that the time spent sorting the data and transmitting it to the SSD does not introduce significant overheads.}

\begin{figure}[t]
         \centering
         \includegraphics[width=0.93\columnwidth]{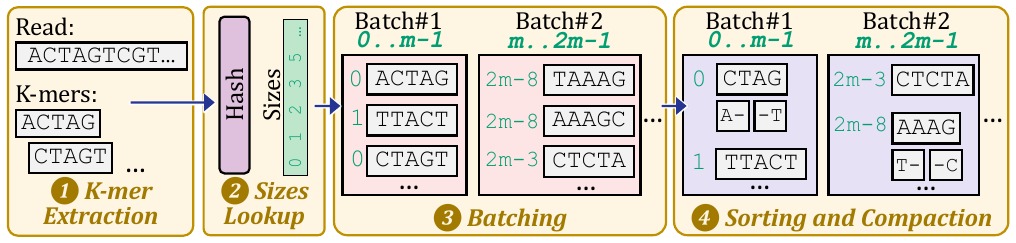}       
  \caption{Overview of \proposal's query processing.}
\label{fig:grains-step1}
\end{figure}

\fig{\ref{fig:grains-step1}} shows an overview of \proposal's query processing. \proposal starts by extracting k-mers from each read (\circled{1} in \fig{\ref{fig:grains-step1}}) and looking them up in \sizes (\circled{2}). To reduce the overhead of sorting and transferring k-mers to the SSD, we take two measures. First, we propose a new k-mer processing scheme by improving upon k-mer processing in \cite{kokot2017kmc3,megis}. We \emph{partition the k-mers into batches corresponding to equal, disjoint ranges of \sizes values} (\circled{3}). Since these values serve as indices into \offsets, this partitioning enables an efficient pipeline: the sorting of one batch overlaps with the data transfer and \offsets lookups of the previous batch. Second, during sorting, we further compact k-mers by exploiting the fact that consecutive k-mers sorted by \sizes share the same minimizer (\circled{4}). Thus, we store the minimizer once and keep only the differences, thereby significantly reducing the transferred data size (e.g., by 2.3$\times$ on average across our evaluated datasets).

\section{Graph Query}
\label{sec:mech-step2}

In this step, \proposal looks up the query k-mers in the graph and, for k-mers that match a graph unitig, retrieves the associated colors. \proposal executes this step inside the SSD 
since it involves accessing large graph structures with low reuse and requires only lightweight computation. This enables us to leverage the SSD’s high internal bandwidth and reduce the burden of moving and analyzing large, low-reuse data on the rest of the system.
In particular, \proposal performs IFP to avoid transferring unnecessary parts of pages out of flash dies and to leverage multiple dies in parallel. 
To efficiently execute this step, \proposal needs to leverage the large internal bandwidth without requiring expensive hardware resources in the SSD (e.g., costly logic units or large  DRAM capacity/bandwidth).
We demonstrate how \proposal meets these requirements when accessing genome graph \offsets (\sect{\ref{sec:mech-offsets}}), \strings (\sect{\ref{sec:mech-strings}}), and \colors (\sect{\ref{sec:mech-colors}}). A lightweight FSM control unit \omiii{in} the SSD controller coordinates execution flow between these stages.

\subsection{Accessing Graph Offsets}
\label{sec:mech-offsets}

\begin{figure}[b]
         \centering
         \vspace{0.1em}
         \includegraphics[width=0.88\columnwidth]{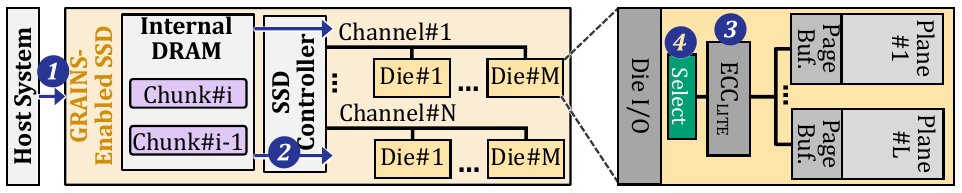}
  \caption{Overview of \proposal{}’s \offsets lookups.}
\label{fig:offsets-lookup}
\end{figure}

\fig{\ref{fig:offsets-lookup}} provides an overview of \proposal{}’s \offsets lookups. \proposal receives \offsets indices and compacted k-mers from the host in chunks (\circled{1}). The internal DRAM holds two small chunks: one incoming from the host and one used for querying \offsets.
For an SSD
with 16 channels, 8 dies/channel, 4 planes/die, and 4-KiB pages,
\proposal needs two 2-MiB chunks.
\proposal reads the values from DRAM, maps them to physical addresses using its simple mapping (\sect{\ref{sec:mech-ds-layout}}), and accesses \offsets (\circled{2}). Since the \sizes values are sorted by the host (\sect{\ref{sec:mech-step1}}) and \offsets are uniformly placed across dies (as detailed in \sect{\ref{sec:mech-ds-layout}}), the lookups lead to sequential accesses that fully exploit the internal bandwidth. 
After loading a page into a die’s page buffer, \proposal applies ECC\textsubscript{LITE}~\cite{lee2025aif} to correct errors (\circled{3}). The IFP units then select and extract only the targeted \offsets entry and send it to the controller (\circled{4}), avoiding the transfer of full pages. \proposal caches the extracted values in internal DRAM for subsequent \strings queries. Although full pages are read inside each die, only a small part of each page is written to internal DRAM, avoiding \omiii{the} internal DRAM bandwidth \omiii{from becoming a} bottleneck.

\subsection{Accessing Graph Strings}
\label{sec:mech-strings}

\omiii{We also aim for the} accesses to the large \strings structure to also fully exploit the SSD’s large internal bandwidth; however, unlike \offsets, the indices into \strings 
(i.e., the retrieved \offsets entries)
arrive unsorted, preventing simple 
streaming.

\head{Lightweight Scheduling} We introduce a lightweight scheduler that \inum{i}~avoids redundant page accesses, and \inum{ii}~maximizes die-level parallelism, without relying on sorting accelerators in the SSD, which are inefficient, given the large number of elements and the limited bandwidth of internal DRAM\cite{zou2022assasin}.

\begin{figure}[t]
         \centering
         \includegraphics[width=0.93\columnwidth]{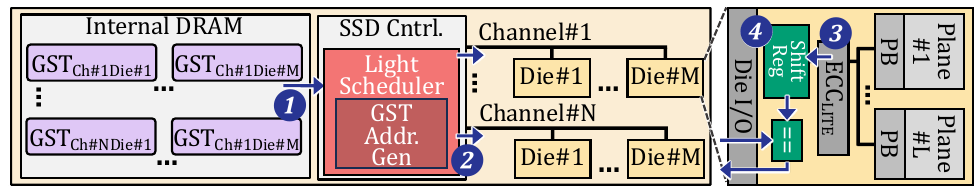}        
  \caption{Overview of \proposal's \strings lookups.}
\label{fig:mech-strings}
\end{figure}

As shown in \fig{\ref{fig:mech-strings}}, \proposal maintains a small per-die table, called \proposal Scheduler Table (\emph{GST}) in the internal DRAM, with one row per \strings page. Each row stores the bit address within the page (as marked by \offsets entries), the k-mer minimizer and its suffixes/prefixes (\fig{\ref{fig:grains-step1}}), a flag indicating whether the row is full (and if so, \omiii{a pointer} to an extension table), and a small one-hot encoded bitmap marking the target plane in the die. The bitmap allows \proposal issue \emph{multi-plane} SSD operations, so that multiple planes of the same die serve different accesses concurrently. 
When accessing \strings, \proposal schedules requests by simply reading one row per GST sequentially and issuing requests across all dies and planes in a round-robin manner. This naturally coalesces accesses to the same page, avoids redundant page accesses, and achieves high parallelism with \omiii{small} metadata and no complex logic.

\proposal can store GSTs in the internal DRAM because \inum{i}~its efficient address mapping (\sect{\ref{sec:mech-ftl-ecc}}) frees most of the DRAM, and \inum{ii}~our compact k-mer representation (\sect{\ref{sec:mech-step1}}) keeps GST entries small. Internal DRAM is sufficient even for large read sets and graphs (e.g., our large evaluated read set with $\sim$10 million query reads on a large-scale genome graph\footnote{\omiii{The graphs we evaluate (\sect{\ref{sec:methodology}}) are among the largest single graphs constructed in practice.}
Graph-based databases that encode even larger volumes of data typically consist of several graphs that need to be queried. This is because graph construction time scales super-linearly, making construction of a single very large graph prohibitively \omiii{slow}~\cite{Bowe2012SuccinctGraphs,pibiri2021pthash}.} requires only 2.9 GB, well within the 4-GB internal DRAM in a typical 4-TB SSD). Furthermore, to ensure efficient functionality even when usage exceeds the internal DRAM's capacity, \proposal keeps some \sizes batches (\sect{\ref{sec:mech-step1}}) in host DRAM until the SSD finishes processing the current batch.

\fig{\ref{fig:mech-strings}} shows an overview of \strings lookups. First, \proposal's scheduler reads \strings addresses and corresponding k-mers from each GST (\circled{1}) and sends the k-mers to the corresponding dies (\circled{2}), incrementing each GST's row pointer for the next access.  Second, \proposal reads the corresponding \strings page, and after ECC\textsubscript{LITE} (\circled{3}), the IFP unit performs a comparison between the \strings \omiii{in} the indexed window and the incoming k-mer from the controller (\circled{4}). 
Finally, \proposal sends the comparison result and the unitig IDs of the matching k-mers back to the SSD controller.

\subsection{Accessing Graph Colors}
\label{sec:mech-colors}

\begin{figure}[h]
         \centering
         \vspace{0.1em}
\includegraphics[width=0.93\columnwidth]{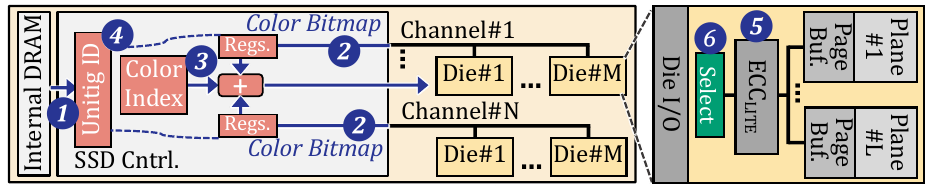}
  \caption{Overview of \proposal's Colors lookups.}
\label{fig:mech-colors}
\end{figure}

For each query k-mer that matches a unitig \omiii{(\sect{\ref{sec:mech-strings}})}, \proposal retrieves the color of that unitig.
Due to \proposal's efficient scheduling (\sect{\ref{sec:mech-strings}}), unitig IDs are already retrieved in the internal DRAM in page order. Thus, given the underlying color encoding (\sect{\ref{sec:mech-ds-layout}}), \omiii{the unitigs'} colors can be obtained via efficient sequential scans of the Color Bitmap (\sect{\ref{sec:mech-ds-layout}}).  \fig{\ref{fig:mech-colors}} illustrates this. \proposal's ISP units stream the next unitig ID from the internal DRAM into a small register (\circled{1}) while concurrently scanning the Colors Bitmap 
from NAND flash dies
(\circled{2}). Given that the bitmap is consumed immediately, \proposal{}'s ISP units directly operate on NAND flash streams (as in prior work~\cite{zou2022assasin}), without buffering them in the internal DRAM. This avoids bottlenecking the internal DRAM's bandwidth.
\proposal uses only two small, 32-bit registers per channel: one for the current and one for the incoming bitmap chunk. As the bitmap is processed, each time a ``\texttt{1}'' is encountered (i.e., a new color is encountered), \proposal increments \emph{Color Index} (\circled{3}). Once the bitmap position matches the unitig ID, \proposal uses the Color Index to index the \colors (\circled{4}). To access \colors, \proposal uses the same approach as \offsets (\sect{\ref{sec:mech-offsets}}), where, after light ECC (\circled{5}), the IFP units select and transmit only the relevant portion of each page (\circled{6}).  The retrieved color entries are returned to the host, where each color identifies the database entries (e.g., species or samples) containing the matched sequence.

\subsection{Detailed Design of the IFP Processing Elements}

\proposal{}'s IFP PEs follow a high-level flow: \inum{i}~a page is read into the die's page buffer, \inum{ii}~the page's content goes through ECC\textsubscript{LITE}, and \inum{iii}~the PE operates on the page buffer contents.

\head{IFP Hardware Units} \proposal{}'s IFP PEs implement two lightweight functions: \inum{i}~\emph{selection}, which extracts a targeted window of a page, and \inum{ii}~\emph{comparison}, which performs a bitwise match between a k-mer and a targeted window. The page buffer in a standard NAND flash die is already organized as a byte- or word-addressable latch-based array with existing sense amplifiers and column select logic. \proposal{}'s selection mechanism repurposes this column decoding circuitry. Given an offset, the column logic isolates the targeted bytes. Instead of streaming the full page over the channel bus, this targeted window is either routed directly to the die's I/O interface or fed internally into the IFP comparison unit. The comparison unit itself consists of a small shift register and a bitwise comparator that checks the k-mer 
against the targeted window.

\head{Interaction with Flash Controllers}
The flash controller sends a small parameter (an offset or a k-mer) to the on-die PE via the existing channel bus. This parameter is stored in a small register on the die and is delivered alongside the standard page read command. After the IFP operations, the controller reads back only the result (instead of the full page).

\section{FTL Integration and Reliability Support}
\label{sec:mech-ftl-ecc}

\proposal FTL requires only minor changes to the baseline FTL. It designates each block as either genomic or non-genomic, and identifies genomic accesses through \proposal commands (\sect{\ref{sec:mech-interface}}). For all non-genomic data, vendor-specific FTL features remain untouched, and the SSD behaves conventionally.  At the start of its SCC operations, since \proposal does not perform writes during them, it flushes the standard L2P metadata to the flash and loads the much smaller \proposal L2P metadata (as detailed below) into internal DRAM, while also keeping the other metadata of a conventional FTL. 

\head{Physical Allocation}
Since \proposal regularizes accesses, \proposal{}'s data placement is simple and enables \inum{i}~efficiently leveraging the SSD's internal bandwidth and \inum{ii}~minimizing mapping metadata. 
We uniformly distribute the blocks of the underlying graph data structures 
and the query reads across different channels, dies, and planes in a round-robin way. For efficient multi-plane operations, active blocks across different planes are aligned to the same page offset.

\head{FTL Metadata Size} 
\proposal reduces the required FTL metadata size due to two reasons. 
First, instead of page-level L2P mappings (which, with 4 KiB pages, leads to $\sim$4~GB of metadata for a 4-TB SSD), \proposal stores L2P metadata at block granularity. This is possible because \proposal{} performs no writes during SCC, and thus, no fine-grained remapping is needed. For a 1-TB graph with 12-MB blocks, this requires only 0.35~MB. Second, \proposal{}'s uniform data placement makes physical locations computable from base addresses and stride, minimizing per-block metadata \omiii{needed for the SSD's reliable operations (e.g., read-disturbance counters)}. With per-block metadata, overall metadata size is 0.7~MB, freeing most of the internal DRAM for \proposal{}'s SCC and scheduling tables.

\head{Error Correction} 
For data accessed by \proposal{}’s ISP units on the SSD controller (e.g., Color Bitmap), accesses occur after the regular ECC. For structures accessed by \proposal's on-die hardware units, we adopt a lightweight on-die ECC technique, \omiii{\emph{ECC\textsubscript{LITE}}}, along with its associated \emph{Bias Error Encoding}~\cite{lee2025aif}.

\head{Other Management Tasks}
Since \omiii{in} SCC mode, \proposal does not incur writes,  it does not require write-related management such as GC and wear-leveling. 
\proposal performs other tasks for ensuring reliability before or after  SCC since 
\inum{i}~the duration of each \proposal process is significantly shorter than the manufacturer-specified threshold for reliable retention age (e.g., a year~\cite{micron3dnandflyer}), and
\inum{ii}~\proposal avoids read-disturbance errors\omiii{~\cite{cai2013error,cai_error_2012,cai2017error,cai2018errorsarxiv,cai-insidessd-2018,micheloni2010inside}} during SCC since, due to its efficient execution, each block is read at most once. After SCC steps, to prevent read disturb\omiii{ance} errors in the future, \proposal refreshes blocks whose read count exceeds a specific threshold.

\head{Profiling-Guided Data Placement} \proposal{} adopts a uniform round-robin data placement across channels, dies, and planes. This data placement works synergistically with \proposal{}'s batching (\sect{\ref{sec:mech-step1}}) and lightweight scheduling (\sect{\ref{sec:mech-step2}}). Specifically, by uniformly distributing data across channels, dies, and planes, round-robin placement ensures that the sorted and batched \offsets queries produced by \proposal{}'s query processing translate directly into well-balanced, parallel accesses across all internal SSD resources. This uniform distribution is also what enables \proposal{}'s lightweight scheduler to \inum{i}~effectively merge accesses to repeated pages and \inum{ii}~efficiently leverage the SSD's internal bandwidth across all dies, without requiring workload-specific profiling. 

That said, for workloads with persistently skewed query distributions (e.g., when specific genomic regions are queried disproportionately often), profile-guided placement could offer additional load balancing benefits across SSD components. Such an optimization would be seamless to integrate with \proposal{}, as it would only affect the initial data layout without requiring changes to \proposal{}'s core batching, scheduling, or ISP/IFP logic. We consider this a promising future work.

\section{Interface Commands}
\label{sec:mech-interface}

\proposal needs three new NVMe commands. 

\texttt{GRNS\_Start} initiates graph genome analysis acceleration. When \omiii{it receives} this command, \proposal prepares itself for \omiii{execution in} SCC mode (\sect{\ref{sec:mech}}). During this phase, \proposal handles execution/data flow based on \proposal FTL and FSM controller.

\texttt{GRNS\_Steps} marks the start and end of each pipeline step and triggers \proposal{}'s lightweight FSM on the SSD controller, which drives the internal execution during SCC. 
The host uses this command to signal that a query batch is ready, then transfers the batch to the SSD's internal DRAM via the standard NVMe data path. Unlike conventional writes, these batches remain in the internal DRAM for processing and are not programmed to flash. The SSD then processes the batch through its internal pipeline with \proposal{}'s FSM on the SSD controller coordinating execution and data flow, and returns results to the host. This enables pipelining multiple batches: as the SSD processes the current batch, the host prepares the next one. No additional commands are needed because the FSM fully determines the internal execution flow once a batch arrives.  At the end of the last step, \proposal returns to operation as a conventional SSD.

\texttt{GRNS\_Write} is a specialized command for writing genome graphs to the SSD, which updates both the regular FTL’s L2P metadata and \proposal{}’s small L2P metadata.

\section{Deployment and Integration}

\proposal{} offers flexible deployment paths to minimize integration costs across both the hardware and software stacks.

\head{Flexible ISP/IFP Deployment} Due to our optimizations, \proposal{}'s ISP/IFP steps require only simple operations with small buffers (\sect{\ref{sec:evals-area}}). Thus, they can execute on either our specialized, lightweight logic units or on general-purpose ISP/IFP (e.g.,~\cite{gu2016biscuit, kang2013enabling, wang2019project,acharya1998active,keeton1998case,riedel1998active,riedel2001active,merrikh2017high,tiwari2013active,tiwari2012reducing,boboila2012active,bae2013intelligent,torabzadehkashi2018compstor,kang2021iceclave,zou2022assasin,chun2022pif,chen2024search}). Specialized logic maximizes power efficiency (\sect{\ref{sec:evals-area}}), while general-purpose ISP/IFP facilitates near-term adoption. Ultimately, choosing between these \proposal{} configurations is a design decision.

\head{Minimal and Non-Disruptive FTL Modifications} The required modifications to the FTL are small and non-disruptive, including simple changes to the baseline FTL, as described in \sect{\ref{sec:mech-ftl-ecc}}. 
The FTL needs to support \proposal{}'s L2P mapping and garbage collection, which are lighter than the baseline (due to \proposal's data placement) and are already modeled in our evaluations by MQSim~\cite{tavakkol2018mqsim,mqsimsource} (\sect{\ref{sec:methodology}}).
Most importantly, when \proposal{} is not operating in its SCC acceleration mode, the SSD operates identically to a standard, general-purpose SSD, which means standard storage deployments are intact.

\head{Seamless System-Level Integration} \proposal{} requires only three new NVMe commands for host-device communication (\sect{\ref{sec:mech-interface}}). This lightweight protocol extension requires no physical interconnect modifications and, similar to prior works with added NVMe commands (e.g.,~\cite{lee2025aif,mansouri2022genstore,megis}), leverages standard vendor-specific NVMe extensions. Similarly, file system and driver modifications are strictly confined to identifying genome graph files and mapping these three commands. This ensures that software integration remains straightforward, low-cost, and non-disruptive to the existing storage stack.

\vspace{-0.3em}
\section{Methodology}
\label{sec:methodology}

\renewcommand\fg{\textsf{FG}\xspace}
\renewcommand\mg{\textsf{MG}\xspace}
\newcommand\pim{\textsf{IdealAccMem}\xspace}
\newcommand\grn{\textsf{GRN}\xspace}
\newcommand\grnext{\textsf{GRN-Ext}\xspace}

\begin{table}[b]
\centering
\scriptsize
\caption{\omiii{Evaluated} SSD configurations.}
\begin{tabular}{@{\hspace{-0.5pt}}c@{\hspace{-0.05pt}}|c|c@{\hspace{-0.05pt}}}
\toprule
\textbf{Specification} & \textbf{SSD-G4} & \textbf{SSD-G5} \\ 
\midrule
\midrule
\textbf{General}       & \multicolumn{2}{c}{\begin{tabular}[c]{@{}c@{}}TLC NAND flash-based SSD\\ 4 TB capacity, 4 GB internal DRAM~\cite{lpddr4} \end{tabular}} \\ 
\midrule
\begin{tabular}[c]{@{}c@{}} \textbf{Bandwidth}\\ \textbf{(BW)} \end{tabular}  & 
\begin{tabular}[c]{@{}c@{}}PCIe Gen4 interface\omiii{~\cite{PCIE4}};\\ 7 GB/s sequential-read BW;\\ 1.2-GB/s channel I/O rate\end{tabular} & 
\begin{tabular}[c]{@{}c@{}}PCIe Gen5 interface\omiii{~\cite{PCIE5}};\\ 14.8 GB/s sequential-read BW;\\ 2.4-GB/s channel I/O rate\end{tabular} \\ 
\midrule
\textbf{NAND Config}  & 
\multicolumn{2}{c}{\begin{tabular}[c]{@{}c@{}}16 channels, 8 dies/channel, 4 planes/die
\end{tabular}} \\ 
\midrule
\bottomrule
\end{tabular}
\label{table:SSD_config}
\end{table}

\head{Performance} 
We design a simulator that models all of \proposal's components, including host operations, internal DRAM, accessing flash dies, in-storage hardware units (on flash dies and on the SSD controller), and SSD-host interfaces. We then feed the latency and throughput of each component to this simulator. Using this methodology, as also leveraged in prior SCC works (e.g.,~\cite{mansouri2022genstore,park2022flash,li2023ecssd,megis,Lee2024presto,mansouri2026sage})\todo{\tiny You recommended citing more. I need to look for more works with this methodology. OK to leave for arxiv?}, enables us to flexibly incorporate state-of-the-art system configurations in our evaluations. For \textbf{hardware-based} \omiii{components} (\omiii{e.g.,} when querying the graph):  We implement \proposal's ISP/IFP logic units in Verilog and synthesize them with a 22nm library~\cite{22gf} using the Synopsys Design Compiler~\cite{synopsysdc}. We use two state-of-the-art simulators, MQSim~\cite{tavakkol2018mqsim,mqsimsource} to model SSD's internal operations, and Ramulator~\cite{kim2016ramulator, ramulatorsource} \omiii{(new version introduced in \cite{luo2023ramulator,ramulator2source})} to model internal DRAM. For \textbf{software-based} \omiii{components} (e.g., when preparing queries), we measure their performance on a real system with 1.5-TB DRAM (in all experiments unless mentioned otherwise) and AMD$^\text{\textregistered}$ EPYC$^\text{\textregistered}$ 7742 CPU~\cite{amdepyc} with 128 physical cores. We use \ssdm~\cite{samsungPM1735} and \ssdh~\cite{samsung9100PRO} as described in \sect{\ref{sec:motivation}} \omiii{and Table~\ref{table:SSD_config}} in our real system experiments. In our MQSim simulations for the ISP/IFP steps, we faithfully model these SSDs with configurations summarized in Table~\ref{table:SSD_config}. For the software baselines, we measure performance using the best-performing thread counts on \omiii{the same} real system \omiii{used for GRAINS's software-based components}. 

To ensure correct composition between host-side and SSD-side timing (including pipelining), we explicitly model the interaction between stages. \proposal{}'s pipeline stages exchange data through well-defined chunk transfers across the host-SSD interface (as described in \sect{\ref{sec:mech-step1}}), which we can faithfully model using the bandwidth and latency of the corresponding interfaces. We enforce stage dependencies via a producer-consumer abstraction; a stage can start only after the required data chunks are produced and transferred, while overlap between stages is naturally limited by the modeled communication bandwidth and device throughput. Doing so enables accurate timing composition of \proposal{}'s lightweight pipeline stages while faithfully accounting for the communication and synchronization costs between host and SSD.

\head{Area and Power}
For \proposal{}'s logic units, we use our Design Compiler synthesis results. 
SSD power is based on values of a Samsung 3D NAND flash-based SSD~\cite{samsung860pro}. DRAM power is based on values from a DDR4 model~\cite{ddr4sheet, ghose2019demystifying}. For the CPU cores, we measure power with AMD \textmu{}Prof~\cite{microprof}.

\head{Evaluated Systems} 
\asp{We evaluate three key tasks: \omiii{\inum{i}}~k-mer set lookup, \omiii{\inum{ii}} alignment-free read mapping, and \omiii{\inum{iii}} alignment-based read mapping. We evaluate the following tools for analysis with large-scale genome graphs:\footnote{\asp{As discussed in \sect{\ref{sec:background-graphs}}, DBGs scale substantially better than other genome graph structures as the genetic diversity in the database increases. Other tools (e.g., VG~\cite{garrison2018vg} and minigraph~\cite{Li2020minigraph}) that use variation graphs as their graph structure do not scale as well and, hence, are not typically used in 
scenarios \omiii{with large-scale, genetically diverse databases}~\cite{karasikov2020metagraph,bvrinda2023efficient}.}}  
\inum{i}~Fulgor~\cite{fan2023fulgor} (\fg), as a state-of-the-art node-centric software tool (which uses SSHash~\cite{pibiri2022sparse}),
\inum{ii}~the MetaGraph framework~\cite{karasikov2020metagraph,karasikov2022lossless,danciu2021topology,karasikov2019sparse} (\mg) as a state-of-the-art edge-centric software tool; 
\inum{iii}~an \emph{idealized} hardware-accelerated baseline (\pim) where all genome graph query operations execute \emph{without} main memory overheads \omiii{(i.e., zero latency, infinite bandwidth, infinite capacity)}. This configuration can represent an \emph{ideal} PIM baseline as well. \pim still incurs the I/O overhead of bringing the large, low-reuse data from the storage system to main memory. Comparing against this baseline demonstrates the impact of storage I/O as a fundamental problem that persists even when main memory overhead is alleviated;
\inum{iv}~\proposal with all its optimizations, but with its lightweight hardware \omiii{as an accelerator} outside the SSD (\grnext), to show the benefits of \proposal{}'s optimizations without ISP/IFP, but with more storage-friendly execution/data flow. The lightweight hardware is connected to the host system via a PCIe interface with 16 GB/s bandwidth, and leverages the host DRAM to store its scheduling metadata (\sect{\ref{sec:mech-strings}}). Similar to \proposal{}'s full implementation, \grnext performs input query processing on the host  (\sect{\ref{sec:mech-step1}}), but instead of relying on its ISP/IFP logic units to query the graph (\sect{\ref{sec:mech-step2}}), it uses the same lightweight \omiii{accelerator} logic units outside the SSD;
\inum{v}~\proposal{}'s full implementation, including ISP and IFP \omiii{(\grn)}.  
All these tools achieve the same accuracy since their underlying graphs are \emph{lossless} encodings of the database's k-mer sets~\cite{karasikov2020metagraph,fan2023fulgor,karasikov2022lossless,danciu2021topology,karasikov2019sparse}. 
For the alignment-based mapping, we integrate all tools with SeGraM, a state-of-the-art sequence-to-graph alignment accelerator~\cite{Cali2022SeGraM}.}

\head{Datasets}
We build graphs as \omiii{described} in \sect{\ref{sec:motivation}}.
The graph (with colors) is 659 GB for \fg, \grn, and \pim, and 822 GB for \mg. We evaluate query read sets with 10M (QL), 1M (QM), and 100K (QS) reads. Note that smaller queries are also critical, as detailed in \sect{\ref{sec:motivation}}.

\section{Evaluation}
\label{sec:evaluation}

\subsection{Performance}

\begin{figure}[h]
    \centering
    \includegraphics[width=\columnwidth]{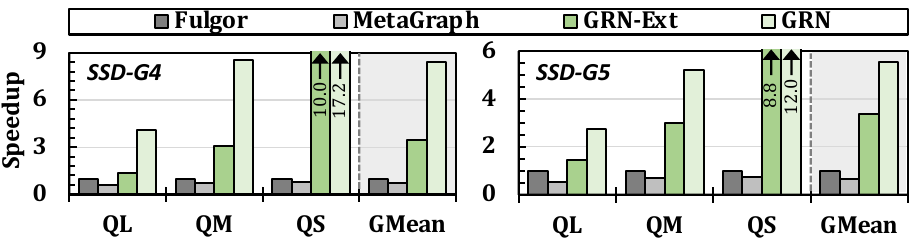}        
    \caption{Speedups with different input query sets and SSDs.}
    \label{fig:grn-eval-main}
\end{figure}

\head{K-mer Set Lookup} \fig{\ref{fig:grn-eval-main}} shows the speedups of different systems over \fg for k-mer set lookups. We make three key observations. First, \proposal provides significant speedups. In systems with \ssdm (\ssdh), \grn provides 8.4$\times$ (5.5$\times$) and 11.9$\times$ (8.5$\times$) average speedups over \fg and \mg, respectively. Second, by making the execution/data flow more storage-friendly, \proposal's implementation outside the SSD also provides large benefits. \grnext provides to 3.4$\times$ and 5.0$\times$ average speedup over \fg and \mg, respectively, across all inputs and SSDs. Third, \proposal's full implementation provides significantly larger benefits over \proposal's implementation outside SSD.
\grn~\omiii{provides}
2$\times$ average speedup over \grnext.
While \grnext exploits locality across queries by merging requests to the same regions of each graph structure and improving access patterns, the locality it captures cannot amortize the cost of transferring the large volume of low-reuse data across the storage-to-host interface. Thus, \proposal{}'s full implementation provides additional, significant benefits by alleviating \omiii{the large} I/O overheads through its SCC \omiii{design}.

\begin{figure}[b]
    \centering
    \includegraphics[width=0.9\columnwidth]{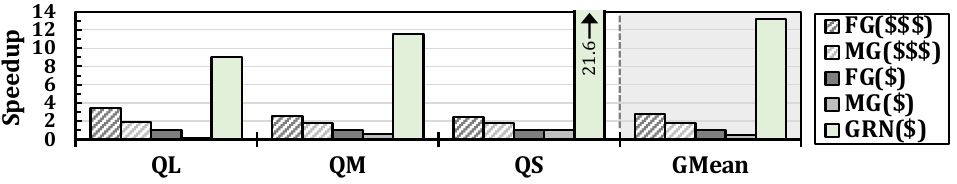}       
    \caption{Speedup of \proposal on a low-cost system, \omiii{i.e., \grn{}(\$)},
    over baselines on low-cost and performance-optimized systems.}
    \vspace{-0.5em}
    \label{fig:grn-eval-cost}
\end{figure}

\head{Impact on Cost Efficiency}
\proposal improves system cost-efficiency since it analyzes large amounts of data in the SSD and does \emph{not} rely on either large host DRAM or high host-SSD interface bandwidth. We show this by evaluating two systems: a cost-optimized system (\$) with 64-GB \omiii{host} DRAM and \ssdm, and a costlier, performance-optimized system (\$\$\$) with 1.5-TB \omiii{host} DRAM and \ssdh. \fig{\ref{fig:grn-eval-cost}} compares \proposal on the cost-optimized system with the baselines on both systems (with a representative query set, QM). To fairly evaluate the baselines when \omiii{host} DRAM is smaller than the graph, we reduce I/O overheads as much as possible in software for this scenario. To this end, 
we partition the graph such that each subgraph fits in the host DRAM, so random accesses to the graph do not repeatedly access the SSD. 
We partition the graph using 
database entry metadata
to guide partitioning choices, 
placing entries 
that likely share many k-mers
into the same partition.
Despite the benefits of this optimization, two sources of overhead remain: \inum{i}~the I/O overhead of transferring subgraphs,
and \inum{ii}~the need to query against each subgraph. 
We make two observations.
First, \proposal on the cost-optimized system significantly outperforms the baselines even on the performance-optimized system. \grn{}(\$) provides 4.7$\times$ and 5.2$\times$ average speedup over \fg{}(\$\$\$) and \mg{}(\$\$\$), respectively. 
Second, the performance of baselines on the cost-optimized system suffers significantly. 
When compared on the same cost-optimized system, \grn{}(\$) provides 13.2$\times$ and 26.9$\times$ average speedup over \fg{} and \mg{}, respectively.\omiiiq{GRN(\$\$\$) was evaluated in the previous figure (as GRN). I will add it to this fig and discussion as well in the arxiv version.} 
We conclude that \proposal improves \omiii{both} system cost-efficiency and \omiii{system} performance, which is critical for facilitating \inum{i}~the wide adoption of graph-based genome analysis, 
and \inum{ii}~analysis on low-cost, portable devices, a scenario rising in importance with the development of portable sequencing devices~\cite{MinIONMk1CO,palatnick2020igenomics,Ballard2018,Oehler2023} for on-site genome analysis~\cite{pomerantz2018real, chiang2019from,mutlu2023accelerating,alser2022molecules}.

\begin{figure}[h]
    \centering
    \includegraphics[width=0.88\columnwidth]{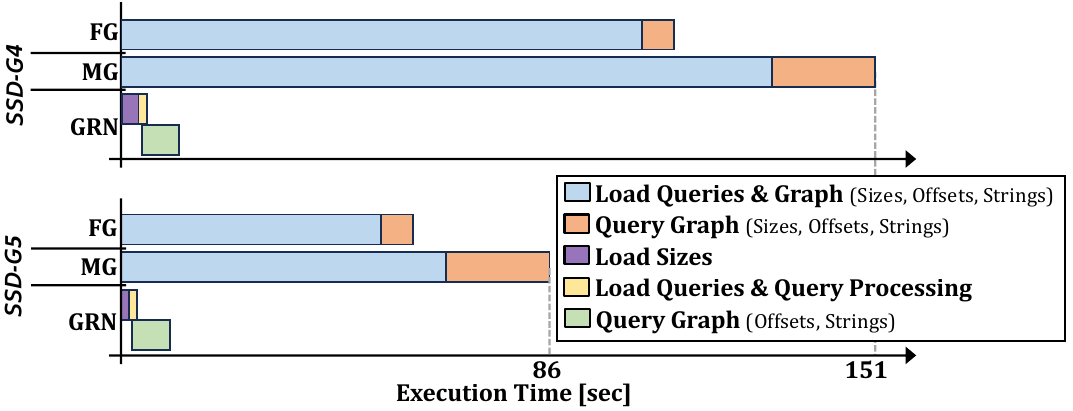}  
    \caption{Execution time breakdown.}
    \label{fig:grn-eval-breakdown}
\end{figure}

\head{Execution Time Breakdown}
\fig{\ref{fig:grn-eval-breakdown}} demonstrates the execution time breakdown of \fg, \mg, and \grn with different SSDs for a representative input query set (QM). In \fg and \mg, the graph and queries need to be first loaded from the storage system to main memory and processing units to be queried. In \grn, first, the \sizes data structure of the graph is loaded from the storage system \omiii{to the host DRAM}. Second, the input query processing step loads the input queries and batches them, as discussed in \sect{\ref{sec:mech-step1}}. Third, as different batches of the input queries are processed, each small batch moves to the storage system in a pipelined manner to query the rest of the graph, without the need to move the data \omiii{(i.e., \offsets, \string, \colors)} outside the SSD, as discussed in \sect{\ref{sec:mech-step2}}. Compared to \fg and \mg, \grn needs to move 31.4$\times$ and 39.1$\times$ less data outside the flash dies, respectively. This \omiii{reduction is due to} \proposal{}'s storage-friendly execution flow, analysis of large-scale graph structures where they originally reside, and efficient scheduling that enables leveraging die-level parallelism.

\newcommand\grnb{\textsf{GRN-B}\xspace}
\newcommand\grnbs{\textsf{GRN-B-S}\xspace}
\newcommand\grnbsscc{\textsf{GRN-B-S-SCC}\xspace}

 \begin{figure}[b]
    \centering
    \includegraphics[width=\columnwidth]{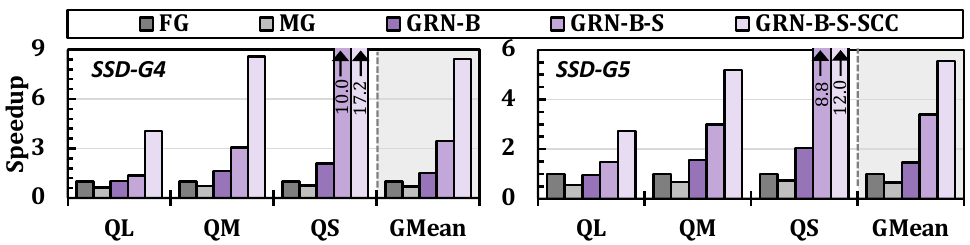}       
    \caption{Ablation tests of different \proposal optimizations.}
    \label{fig:grn-eval-ablation}
\end{figure}

\head{Effect of Different \proposal Optimizations} \fig{\ref{fig:grn-eval-ablation}} demonstrates the benefits of different \proposal optimizations. \grnb is a \proposal configuration including only the batching optimization (\sect{\ref{sec:mech-step1}}). \grnbs is a \proposal configuration including both the batching and scheduling (\sect{\ref{sec:mech-strings}}) optimizations (equivalent to \grnext, \sect{\ref{sec:methodology}}). \grnbsscc is \proposal{}’s full configuration, including its SCC. Please note that we do not show a configuration with naive SCC and without batching and scheduling. This is because simply using SCC without these techniques leads to performance loss due to the irregular access patterns to genome graphs, 
\omiii{which are inefficient}
due limitations of the SSD’s hardware \omiii{(e.g., costly conflicts in internal SSD resources such as channels and NAND flash chips~\cite{nadig2023venice,tavakkol2018flin,kim2022networked} during random and dependent accesses)}. We make three observations. First, \grnb provides 1.5$\times$ and 2.2$\times$ average speedup over \fg and \mg, respectively. This is due to more efficient accesses to the \offsets data structure. Second, \grnbs provides 2.3$\times$ average speedup over \grnb as it enables more storage-friendly accesses to \strings and \colors as well. Finally, through its efficient SCC, \grnbsscc provides 2.0$\times$ and 4.6$\times$ average speedup over \grnbs and \grnb, respectively.

\begin{figure}[h]
    \centering
    \includegraphics[width=\columnwidth]{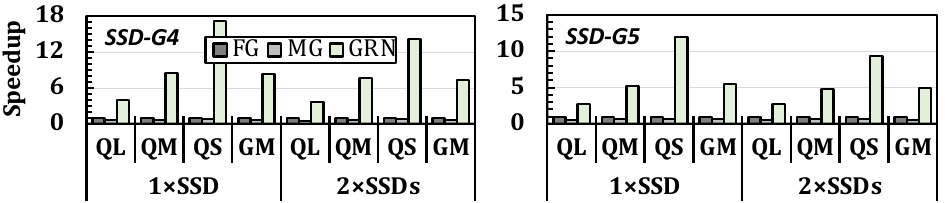}       
    \caption{Speedups with different numbers of SSDs.}
    \label{fig:grn-eval-multi-ssd}
\end{figure}

\head{Effect of the Number of SSDs}
\proposal can benefit from multiple SSDs in two ways: First, different graphs can be stored on separate SSDs and analyzed concurrently, each benefiting from \proposal (as shown in \fig{\ref{fig:grn-eval-main}}). 
Second, a single graph can be partitioned across SSDs to increase throughput. This is enabled by \proposal{}’s execution flow supporting efficient graph partitioning.
Since \proposal manages \offsets indices in the host, \offsets can be divided disjointly, each query being directed to the relevant SSD. \strings and \colors can also be partitioned: based on \offsets values, \proposal determines whether the referenced \strings entry resides locally or remotely.  
If remote, \proposal forwards values to the corresponding SSD. Given that \proposal{}'s IFP only selects and transfers relevant parts of \offsets, \proposal avoids read amplification. To further minimize communication overhead, \proposal batches \offsets values and maintains two 2-MB batches in the internal DRAM's freed-up space (\sect{\ref{sec:mech-ds-layout}}) so that while one batch is transmitted, the other can be filled, thereby overlapping communication with SCC and efficiently utilizing internal and external bandwidths.
We evaluate this scenario in \fig{\ref{fig:grn-eval-multi-ssd}}. We observe that 
\proposal maintains large speedups. This is because, as the external bandwidth increases for the baselines with more SSDs, the internal bandwidth also increases.
Although there is a slight decrease in speedup when moving to two SSDs in QS, the speedup is still significant (14.2$\times$ and 9.4$\times$ with \ssdm and \ssdh). This decrease is because, after increasing internal bandwidth, the impact of software query processing becomes relatively larger. Therefore, we can further accelerate these steps (\omiii{e.g.}, with a sorting accelerator) to increase benefits even further. 
We conclude that  \proposal can efficiently leverage multiple SSDs, making it also suitable for distributed systems.\omiiiq{We will add more SSDs for the extended version}

\begin{figure}[b]
    \centering
    \includegraphics[width=\columnwidth]{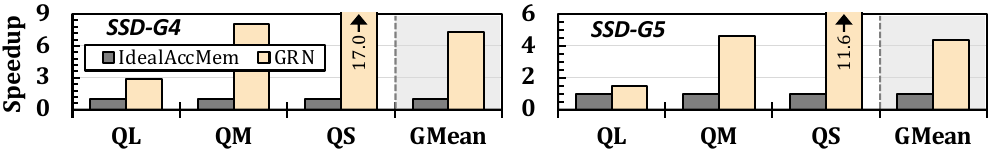}       
    \caption{Speedup over IdealAccMem.}
    \label{fig:grn-eval-pim}
\end{figure}

\head{Comparison to IdealAccMem}  \fig{\ref{fig:grn-eval-pim}} shows speedups over \pim. Despite \pim{}'s benefits, it still incurs I/O accesses to transfer data to main memory, even when DRAM is larger than data size (as in our evaluation). 
Through its storage-aware design, with efficient ISP/IFP operations, we observe that on the system with \ssdm (\ssdh), \proposal provides 7.4$\times$ (4.3$\times$) average speedup over \pim.

\begin{figure}[h]
    \centering
    \includegraphics[width=\columnwidth]{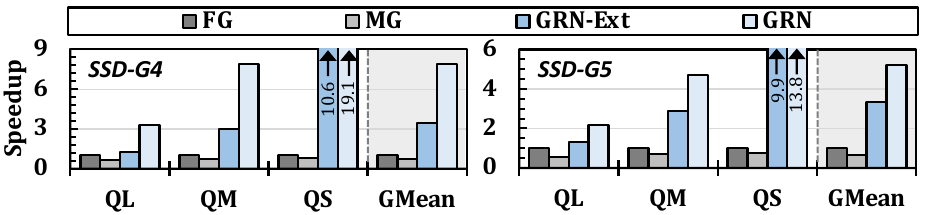}       
    \caption{Speedups for read mapping.}
    \label{fig:grn-eval-mapping}
\end{figure}

\head{\asp{Alignment-Free} \omiii{Read} Mapping} \fig{\ref{fig:grn-eval-mapping}}  shows speedups over \fg when performing \omiii{alignment-free} read mapping. \omiii{In this case, as discussed in \sect{\ref{sec:mech-step1}}, \proposal sends the colors of all k-mers to the host system, and the host system determines the colors of each read based on the colors of its k-mers.} We observe that on systems with \ssdm (\ssdh), \grn provides 7.9$\times$ (5.2$\times$) and 11.2$\times$ (8.0$\times$) average speedups over \fg and \mg, respectively. 

\begin{figure}[ht]
    \centering
    \includegraphics[width=\columnwidth]{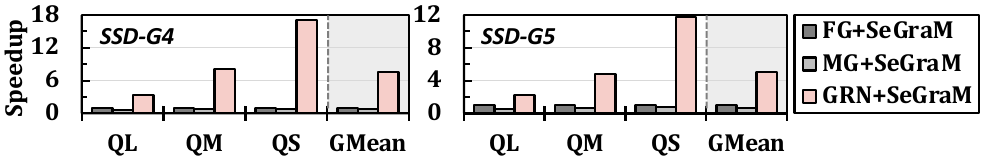}       
    \caption{Speedups for read mapping with alignment.}
    \label{fig:grn-eval-alignment}
    \vspace{-0.2em}
\end{figure}

\head{\asp{Alignment-Based Mapping}}
We integrate \grn, \fg, and \mg with SeGraM~\cite{Cali2022SeGraM} to perform alignment on the candidate graph regions identified by each tool (we use the throughput of alignment operations reported by \omiii{SeGraM}~\cite{Cali2022SeGraM}). \fig{\ref{fig:grn-eval-alignment}} shows the speedups of different systems over \fg+SeGraM.
We observe that SeGraM+\grn provides 6.2$\times$ and 9.0$\times$ average speedup over SeGraM+\fg and SeGraM+\mg. 

\subsection{Area and Power \omiii{Overheads}}
\label{sec:evals-area}
Table~\ref{table:area-power} summarizes area and power for \proposal{}'s logic units at 333MHz. While these units can operate at higher frequencies, their throughput is already sufficient since \proposal's ISP/IFP pipelines are bottlenecked by flash reads. We use area/power values of ECC\textsubscript{LITE} from the original work~\cite{lee2025aif}.\footnote{Since we do not have access to the larger technology node used in \cite{lee2025aif}, we scale ECC\textsubscript{LITE}'s area to the lower node with the methodology in \cite{stillmaker2017Scaling}.}
\proposal{}'s hardware overhead is very small. On-controller units are 0.0025 mm\textsuperscript{2} (0.7\% of the four ARM cores~\cite{cortexr4} on an SSD controller). \proposal's on-die logic area is mainly (99\%) for ECC\textsubscript{LITE}, which, as reported by \cite{lee2025aif}, takes only 0.2\% of the flash chip area and fits well within its power budget.

\begin{table}[b]
\centering
\vspace{-0.2em}
\caption{Area and power \omiii{consumption} of \proposal's logic units.}
\label{tab:metastore_area_energy}
\resizebox{0.67\columnwidth}{!}{%
\begin{tabular}{c|c|c}
\toprule
\textbf{Logic unit}                  & \textbf{Area [mm\textsuperscript{2}]} & \textbf{Power [mW]} \\ 
\midrule
\midrule
On SSD Controller                  &    0.0025    &      0.21    \\
On Die (ECC\textsubscript{LITE})   &    0.036    &      18.04    \\
On Die (Others)                    &    0.000093    &      0.01    \\\bottomrule
\end{tabular}
}
\label{table:area-power}
\end{table}

\asp{As detailed in \sect{\ref{sec:mech-step2}}, \proposal{}’s SCC can execute on either \inum{i}~our lightweight, specialized ISP/IFP units or \inum{ii}~general-purpose ISP (e.g.,~\cite{gu2016biscuit, kang2013enabling, wang2019project,acharya1998active,keeton1998case,riedel1998active,riedel2001active,merrikh2017high,tiwari2013active,tiwari2012reducing,boboila2012active,bae2013intelligent,torabzadehkashi2018compstor,kang2021iceclave,zou2022assasin}) and IFP (e.g.,~\cite{chun2022pif,chen2024search}). 
Choosing between these \proposal{} configurations is a design tradeoff: general\omiii{-purpose} units facilitate earlier adoption, while specialized units offer better area and power efficiency. For example, \proposal{}'s ISP units on the SSD controller consume two orders of magnitude lower area and 60.1$\times$ lower power than hardware units of a state-of-the-art general\omiii{-purpose} ISP.\footnote{\asp{Note that the power/area of \proposal{}'s IFP units are mainly dominated by ECC\textsubscript{LITE}, needed in both  \proposal and other IFP systems (e.g.,~\cite{chun2022pif, lee2025aif}).}}}

\subsection{Energy} 
We find \omiii{the energy consumption of each evaluated graph-based genome analysis tool (when performing k-mer set lookups)} based on \omiii{the energy consumption of} different components (e.g., host processor, host DRAM, host-SSD communication, SSD components, and logic units). We calculate each component's energy based on its dynamic/idle power and execution time. 
We observe that across our evaluated input queries and SSDs, \proposal provides 4.4--21.0$\times$, 12.2--31.6$\times$, and 3.1--20.7$\times$ energy reduction over \fg, \mg, and \pim, respectively, when performing k-mer set lookups.\omiiiq{You mentioned the energy results are side-lined. I will add a figure to add more emphasis on them in the extended version. Please let me know if you suggest more actions.}

\section{Related Work}
\label{sec:related}

To our knowledge, GRAINS is the first \omiii{storage-centric computing (SCC)} system designed to significantly reduce I/O data movement overheads and improve end-to-end performance, energy-efficiency, \omiii{and cost-effectiveness} of graph-based genome analysis.

\head{Analysis with Genome Graphs}
Many works (e.g., \cite{rautiainen2020graphaligner,kim2019hisat2,gao2020abpoa,jain2019pasgal,siren2021pangenomics,Rautiainen2019,Chandra2023,Ivanov2022,Ma2023,Darby2020vargas,Hwang2025MEMO,Romain2023svjedi,Li2020minigraph}) 
propose software tools for genome graphs. Several works propose hardware tools for querying genome graphs by alleviating computation (e.g.,~\cite{Cali2022SeGraM,Zhang2024Harp,Zeng2024asgdp,Shen2024128parallel,Li2024,Mandal2020,Varma2013,Awan2021,Feng2021,Zhang2025}) and/or memory overheads (e.g.,~\cite{Cali2022SeGraM,kim2025nmp,Huang2023meg2,Angizi2020Panda,Qiu2017}). However, these works do not alleviate storage I/O overheads, whose impact on end-to-end performance becomes even larger as other overheads are alleviated. As shown in \fig{\ref{fig:grn-eval-alignment}}, \proposal can flexibly integrate with these designs to alleviate their I/O overhead.
Some works (e.g.,~\cite{Zhou2021,Sarkar2021,Varma2017,Varma2016,Goswami2018,Galanos2021,Angizi2020,Sinha2022,Meng2014,Hu2016,Chen2023,Natarajan2018,Ren2018}) accelerate genome graph \emph{construction}, which is an important, but orthogonal task.

\head{Accelerating Graph Analysis}
Many works accelerate graph analysis by alleviating computation (e.g.,~\cite{Dadu2021polygraph, Ham2016Graphicionado,Rahman2020graphpulse,Song2018graphr,Chen2022regraph,Yang2025IDGNN,Yan2025bingogcn,Peng2024maxkgnn,Yan2020hygcn,You2022gcod,Hwang2023grow,Sarkar2023flowgnn,Chen2022regnn,Li2021gcnax,Geng2020awbgcn,Chen2023metanmp,Zhao2025mehyper}), main memory (e.g.,\omiii{~\cite{onur2015dramgraph,Zhou2022GNNear,zhang2018graphp,Asiatici2021,Shin2025piccolo,Huang2022reflip,Besta2021SISA,Chen2023metanmp,Li2022hyperscale,Wang2024motionaccel,Dai2023cegma,Kim2025eod,Li2024celeritas,ahn_pim-enabled_2015,kanellopoulos_smash_2019,dai2018graphh}}), or I/O (e.g.,~\cite{matam2019graphssd,Wang2024ndsearch,lee2022smartsage,Niu2024flashgnn,Lee2024presto,Khadirsharbiyani2024smartgraph,Zhang2025taijigraph,An2023baraddur,Kang2024sting}) overheads.
However, they are neither sufficient for nor target the unique demands of genome graphs (as detailed in \sect{\ref{sec:graph-uniqueness}}).

\head{Accelerating Genome Analysis}
Many works 
(e.g.,\omiii{~\cite{wang2023gpmeta, Gamaarachchi2020_f5c, ahmed2019gasal2, fujiki2020seedex, rucci2018swifold, ham2020genesis, liyanage2023efficient, turakhia2018darwin, fujiki2018genax,wu2021sieve, dashcam23micro,zou2022biohd, cali2020genasm, mansouri2022genstore, megis, Walia2024talco,Huangfu2022beacon, Huangfu2019medal,Lindegger2023scrooge, Diab2023, diab_framework_2022,Sun2025, Barkam2024, Simon2025cimba, dong2024mm2, DeMoor2024mimycs, Wu202528nmendtoend,alser2020accelerating,singh2021fpga,alser2022molecules,mutlu2023accelerating,simon2026pim,shahroodi2023swordfish,shahroodi2022demeter,kim2018grim,singh2024rubicon,eudine2026genpairx,Pavon2024quetzal,alser2020sneakysnake,alser2017gatekeeper,bingol2021gatekeeper,thesis,alser2019shouji,alonso2024bimsa,firtina2024aphmm,firtina2025enabling,senol2021accelerating,koliogeorgi2023hardware,Doblas2023gmx,doblas2025smx,nag2019gencache,lou2020helix,lou2018brawl,markus2020benchmarking,subramaniyan2021accelerated,huangfu2018radar,khatamifard2021genvom,gupta2019rapid,li2021pim,angizi2019aligns,zokaee2018aligner,madhavan2014race,cheng2018bitmapper2,houtgast2018hardware,houtgast2017efficient,zeni2020logan,nishimura2017accelerating,de2016cudalign,liu2015gswabe,liu2013cudasw++,liu2009cudasw++,liu2010cudasw++,wilton2015arioc,goyal2017ultra,chen2016spark,chen2014accelerating,chen2021high,banerjee2018asap,fei2018fpgasw,waidyasooriya2015hardware,chen2015novel,haghi2021fpga,li2021pipebsw,ham2021accelerating,wu2019fpga,Zhang_2023_alignerD,kaplan2020bioseal,mao2022genpip,dphls2026,wang20202,sadasivan2024genomic,Turakhia2025toward,Turakhia2019darwinwga}}) propose hardware optimizations for conventional genome analysis on linear sequences, including \omiii{SCC} systems (e.g.,~\cite{mansouri2022genstore,abakus23taco,megis,jun2016storage,kim2025nmp,soysal2025mars,Zheng2025ispgenome,tsai2025accelerating}) for these analyses. However, none of them are suitable for graph-based genome analyses due to fundamentally different data structures, execution flow, and irregular and dependent access patterns.

Particularly, \proposal{} \omiii{exploits the properties of graph-based genome analyses in three ways.} 
First, since \proposal{} needs to handle \emph{dependent accesses} across \emph{large data structures residing in the SSD}, it needs to handle scheduling directly in the SSD. To do so, \proposal{} introduces a novel data/execution flow and lightweight scheduling (\sect{\ref{sec:mech-step2}}), fundamentally different from prior approaches that regularize accesses purely through query scheduling in the host~\cite{megis} or via offline preprocessing~\cite{mansouri2022genstore}. Second, \proposal{}'s batching and task partitioning \omiii{efficiently exploit} the underlying properties of genome graphs (\sect{\ref{sec:mech-step1}}), \omiii{which no prior work considers}. Third, \proposal{} performs end-to-end sequence-to-graph read mapping (which is more complex than traditional sequence-to-sequence mapping).

Several works propose techniques for batching k-mer queries to improve access patterns to genomic data structures (e.g.,~\cite{alanko2025batched, zhang2013optimizing, li2024high}). Despite their benefits, these techniques cannot be directly adopted in SCC designs for genome graphs due to the dependent and random accesses to different data structures.
A purely host-side approach would require a separate round trip to the SSD for each dependent access stage, \omiii{undermining} the locality benefits of batching. On the other hand, directly adopting these techniques within the SSD stresses the limited available hardware resources in modern SSDs.
\proposal{} addresses \omiii{these limitations} with a co-designed approach. First, we perform lightweight, genome-graph-aware batching on the host (exploiting the small \sizes metadata in the host, \sect{\ref{sec:mech-step1}}). Second, once these batched requests are dispatched to the SSD, our new, lightweight in-storage scheduling orchestrates all subsequent dependent accesses to large data structures inside the SSD, without \omiii{any} further round-trips to the host (\sect{\ref{sec:mech-step2}}).

\head{SCC Designs}
Various works propose SCC 
for different applications (e.g., in AI/ML \cite{lee2025aif,li2023ecssd,Adnan2024,Peng2025xharvest,Tang2024transformerisp,liang2019ins,lee2022smartsage,wang2024beacongnn,jang2024smart,Chen2025reis,Yu2024cambriconllm,Niu2024flashgnn,Kim2023optimstore,Sun2025lincoln,Taranco2025iris,Pan2025instattention,mailthody2019deepstore}, \omiii{string matching} and read mapping\omiii{~\cite{jun2016storage,mansouri2022genstore,Hsu2024het3dnand,Zheng2025ispgenome,soysal2025mars}}, k-mer counting \cite{abakus23taco}, graph analytics~\cite{jun2018grafboost,xu2019vstore,matam2019graphssd}, and others~\cite{Mahapatra2025isp_rag,liang2019cognitive,kim2020reducing,lim2021lsm,li2021glist,wang2016ssd,lee2020neuromorphic,kang2021s,han2021flash,wang2022memcore,wang2018three,han2019novel,choi2020flash,pei2019registor,do2013query,seshadri2014willow,kim2016storage,jeong2019react,Duffy2023dotori,Park2025anykey,Chen2024aresflash,Wong2025anvil,mahapatra2024instoragedomainspecificaccelerationserverless}). Several works propose general-purpose SCC in storage\omiii{~\cite{gu2016biscuit, kang2013enabling, wang2019project,acharya1998active,keeton1998case,riedel1998active,riedel2001active,merrikh2017high,tiwari2013active,tiwari2012reducing,boboila2012active,bae2013intelligent,torabzadehkashi2018compstor,kang2021iceclave,zou2022assasin,nadig2026conduit,mansouri2026sage,mansouri2026sagearxiv}}, SSDs integrated closely with FPGAs~\cite{jun2015bluedbm, jun2016bluedbm, torabzadehkashi2019catalina, lee2020smartssd, ajdari2019cidr, koo2017summarizer,Jeong2025upp}, or with GPUs~\cite{cho2013xsd}. None of these works targets analysis with genome graphs.

Some works (e.g.,~\cite{kang2022pr,lee2025srnand}) propose performing sub-page reads to avoid unnecessary data movement from the flash dies. However, they focus solely on reducing data transfer granularity and do not perform computation within the dies.
Several works propose performing computation
in the dies~\cite{chun2022pif,chen2024search,lee2025aif}, or using flash memory\omiii{~\cite{gao2021parabit,park2022flash,Chun2024rif,Chen2024aresflash}}. However, directly applying these techniques to genome graph analysis is \omiii{undesirable} due to the irregular, data-dependent accesses, which hinders leveraging die-level parallelism. \proposal{} addresses this challenge through its genome-graph-aware \omii{algorithm-architecture} co-design.

Several works (e.g.,~\cite{Kim2023optimstore, nadig2026conduit, liang2024hyqa}) propose SCC designs that combine ISP, IFP, and PIM. However, these works are not sufficient to handle the unique properties of graph-based genome analysis, which pose challenges absent from prior SCC work. 
We address these challenges via our genome-graph-aware query reordering techniques and lightweight scheduling, enabled via \proposal{}'s holistic co-design of data/execution flow, data layout, and FTL integration.

\section{Conclusion}
\label{sec:conclusion}

We introduced \proposal, the first storage-centric system designed to alleviate the I/O overhead of analysis with large genome graphs. Through detailed examination of these analysis pipelines, we \inum{i}~make them more
storage-friendly and \inum{ii}~enable efficient in-storage and in-flash processing. Our evaluations show that \proposal improves performance, energy efficiency, and system cost-effectiveness of graph-based genome analysis at low area and power costs.
We hope that \proposal{}'s optimizations will inspire and benefit \omiii{storage-centric computing} systems in other domains to likewise improve their performance and energy efficiency without relying on expensive hardware resources.

\section*{Acknowledgments}

We thank the anonymous reviewers of ASPLOS 2026 and ISCA 2026 for feedback. We thank the SAFARI group members for feedback and the stimulating, inclusive, intellectual, and scientific environment. We acknowledge the generous gifts and support provided by our industrial partners, including Google, Huawei, Intel, Microsoft, and VMware. This research was partially supported by the European Union’s Horizon Program for research and innovation under Grant  No. 101047160 (project BioPIM), the Swiss National Science Foundation (SNSF), Semiconductor Research Corporation (SRC), the ETH Future Computing Laboratory (EFCL), Huawei ZRC Storage Team, and the AI Chip Center for Emerging Smart Systems Limited (ACCESS). Jisung Park was supported by the NRF (RS-2025-00519994) and the IITP (RS-2024-00459026) of Korea.

\bibliographystyle{unsrt}
\bibliography{refs}

\end{document}